\documentclass[11pt]{article}
\def\llsymbol#1{\@llsymbol{\@nameuse{c@#1}}}
\def\@llsymbol#1{\ifcase#1\or {}\or {'}\or {''}\or {'''}\or
ÃÂ  ÃÂ {''''}\or {'''''}\or ÃÂ \else\@ctrerr\fi\relaz}
\newcounter{contador}
\newcommand{\letra}{
ÃÂ  ÃÂ \stepcounter{equation}
ÃÂ  ÃÂ \setcounter{contador}{\value{equation}}
ÃÂ  ÃÂ \setcounter{equation}{0}
ÃÂ  ÃÂ \renewcommand{\theequation}{\thecontador\alph{equation}}}
\newcommand{\antiletra}{
ÃÂ  ÃÂ \renewcommand{\theequation}{\arabic{equation}}
ÃÂ  ÃÂ \setcounter{equation}{\value{contador}}}

\usepackage{indentfirst}
\usepackage{eucal}
\usepackage{amsmath}
\usepackage{amsfonts}
\usepackage{amssymb}
\usepackage{mathrsfs}
\begin{document}
\begin{center}{ { \bf ON CERTAIN SOLUTIONS FOR CONFLUENT AND DOUBLE-CONFLUENT HEUN EQUATIONS
}\\
\vspace{.6cm}
L\'ea Jaccoud El-Jaick and Bartolomeu D. B. Figueiredo}\\
{\small Centro Brasileiro de Pesquisas F\'{\i}sicas (CBPF)\\
Rua Dr. Xavier Sigaud, 150, CEP 22290-180, Rio de Janeiro, RJ, Brasil}
\end{center}
%

%
%

%
\noindent
{\bf Abstract}. This paper examines some solutions for
confluent and double-confluent
Heun equations. In the first place, we review
two Leaver's solutions
in series of regular and irregular confluent
hypergeometric functions for the confluent
equation 
and  introduce an additional
expansion in series of irregular confluent hypergeometric
functions. Then, we find the conditions under which
one of these solutions can be written as a linear
combination of the others.
In the second place, by means of
limiting  procedures we generate solutions
for the double-confluent equation as well as for
special limits of both the confluent and double-confluent
equations.  Finally, we present problems which
are ruled by each of these four
equations and establish relations among
Heun equations and quasi-exactly solvable problems.
%

\section*{1. Introduction}

This paper is concerned with solutions of Heun equations
and possible applications of such solutions.
We consider only the confluent  and the double-confluent
Heun equations (CHE and DCHE, respectively), and one
limiting case of each of these.
The solutions for the CHE come directly from
the differential equation, while the solutions for the other equations
are obtained from the solutions of the CHE by limiting processes.
Initially, we briefly discuss each of these
equations and their connections;
some more details are found in
previous works \cite{eu1,eu2,eu3,leaver1}.
Then, we outline the main features of
the solutions  and  the
structure of the paper.

The CHE \cite{decarreau1,decarreau2,ronveaux}, also known as generalised
spheroidal wave equation \cite{wilson1,wilson2}, in the form used by
Leaver \cite{leaver1} reads ($\omega\neq0$)
\begin{eqnarray}
\label{gswe}
z(z-z_{0})\frac{d^{2}U}{dz^{2}}+(B_{1}+B_{2}z)
\frac{dU}{dz}+
\left[B_{3}-2\eta
\omega(z-z_{0})+\omega^{2}z(z-z_{0})\right]U=0,
\end{eqnarray}
where $B_{i}$, $\eta$ and $\omega$ are constants
and $z=0$ and $z=z_{0}$ are regular singular
points with indicial exponents ($0,1+B_{1}/z_{0}$)
and ($0,1-B_{2}-B_{1}/z_{0}$), respectively.
At the irregular point $z=\infty$ the
behaviour of the solutions, obtained from the
normal Thom\'e solutions \cite{leaver1,olver}, is given by
\begin{eqnarray}\label{thome1}
\lim_{z\rightarrow ÃÂ \infty}U(z)\sim e^{\pm i\omega z}z^{\mp i\eta-(B_{2}/2)}.
\end{eqnarray}
%
%

The singularity parameter $z_{0}$
may take any value and, when $z_{0}=0$,  the CHE
gives the following DCHE with five parameters \cite{leaver1}
\begin{eqnarray}\label{dche}
z^{2}\frac{d^{2}U}{dz^{2}}+
\left(B_{1}+B_{2}z\right)\frac{dU}{dz}+
\left(B_{3}-2\eta \omega z+\omega^{2}z^{2}\right)U=0,
\ \left(B_{1}\neq 0, \ ÃÂ \omega\neq 0\right),
\end{eqnarray}
where now $z=0$ and $z=\infty$ are both irregular
singularities ($B_1=0$ and/or $\omega=0$ are degenerate
cases \cite{eu2}). At $z=\infty$ the behaviour is again
given by Eq. (\ref{thome1}), while at $z=0$ the normal
Thom\'e solutions afford
\begin{eqnarray}\label{thome}
\displaystyle\lim_{z\rightarrow ÃÂ 0}U(z)\sim 1,\ \mbox{or}\
ÃÂ \displaystyle\lim_{z\rightarrow ÃÂ 0}
U(z)\sim e^{B_{1}/ z}z^{2-B_{2}}.
\end{eqnarray}

The CHE and the DCHE admit a limit which changes
the nature of the irregular singularity at $z=\infty$,
keeping unaltered the other singular points.
This limit is obtained by letting that \cite{eu2,eu3}
\begin{eqnarray}\label{ince}
\omega\rightarrow 0, \ \
\eta\rightarrow
\infty, \ \mbox{such that }\ ÃÂ \ 2\eta \omega =-q,\ \
(\mbox{Whittaker-Ince limit})
\end{eqnarray}
where $q$ is a constant. It is called Whittaker-Ince
limit because Whittaker and Ince have used a similar procedure
to get the Mathieu equation (\ref{mathieu}) from
the Whittaker-Hill equation (\ref{whe}) \cite{humbert,ince}.
The Whittaker-Ince limit of the CHE is
\begin{eqnarray}
\label{incegswe}
z(z-z_{0})\frac{d^{2}U}{dz^{2}}+(B_{1}+B_{2}z)
\frac{dU}{dz}+
\left[B_{3}+q(z-z_{0})\right]U=0,\qquad (q\neq0)
\end{eqnarray}
(if $q=0$ this equation can be transformed into a hypergeometric
equation), while the Whittaker-Ince limit of the DCHE is
\begin{eqnarray}
\label{incedche}
z^2\frac{d^{2}U}{dz^{2}}+(B_{1}+B_{2}z)
\frac{dU}{dz}+
\left(B_{3}+qz\right)U=0,\qquad (q\neq0, \ B_{1}\neq 0)
\end{eqnarray}
(if $q=0$ and/or $B_1=0$
the equation degenerates into a confluent hypergeometric
equation or simpler equations \cite{eu2}).
Eqs. (\ref{incegswe}) and (\ref{incedche})
differ from the CHE and DCHE, respectively,
by the behaviour of their solutions
at the irregular point $z=\infty$, which now is obtained
from the subnormal Thom\'e solutions \cite{olver}, namely,
\begin{eqnarray}\label{thome2}
\lim_{z\rightarrow ÃÂ \infty}U(z)\sim
e^{\pm 2i\sqrt{qz}}z^{(1/4)-(B_{2}/2)},
\end{eqnarray}
in contrast with the behaviour of original equations
(normal Thom\'e solutions).
Eq. (\ref{incedche}) also results when we take $z_0=0$
in Eq. (\ref{incegswe}).

The preceding equations
and their connections are summarised in
the following diagram which is a modified version
of a diagram given in \cite{eu3}. The upper  boxes
display the CHE and the DCHE. The lower
boxes  show the
Whittaker-Ince limits corresponding to the CHE and DCHE,
respectively.
\begin{eqnarray}
\boxed{
\begin{array}{l}
\displaystyle z(z-z_{0})\frac{d^{2}U}{dz^{2}}+(B_{1}+B_{2}z)
\frac{dU}{dz}+\\
\\
\left[B_{3}-2\eta
\omega(z-z_{0})+\omega^{2}z(z-z_{0})\right]U=0
\end{array}} \stackrel{z_0\rightarrow0}{\Rightarrow}
\boxed{
\begin{array}{l} \displaystyle z^{2}\frac{d^{2}U}{dz^{2}}+
\left(B_{1}+B_{2}z\right)\frac{dU}{dz}+\\
\\
\left(B_{3}-2\eta \omega z+\omega^{2}z^{2}\right)U=0
\end{array}}
\end{eqnarray}
\begin{eqnarray*}
\Downarrow\quad(\omega\rightarrow 0\ \text{and}\
\eta\rightarrow \infty,\  \text{such that }
2\eta\omega=-q)\qquad\Downarrow
\end{eqnarray*}
\begin{eqnarray}
\boxed{
\begin{array}{l} \displaystyle \
z(z-z_{0})\frac{d^{2}U}{dz^{2}}+(B_{1}+B_{2}z)
\frac{dU}{dz}+\hspace{1.05cm}\\
\\
\ \left[B_{3}+q
(z-z_{0})\right]U=0
\end{array}} \stackrel{z_0\rightarrow0}{\Rightarrow}
\boxed{
\begin{array}{l}\displaystyle
z^{2}\frac{d^{2}U}{dz^{2}}+
\left(B_{1}+B_{2}z\right)\frac{dU}{dz}+\\
\\
\left(B_{3}+qz\right)U=0
\end{array}}
\end{eqnarray}
These connections among  equations
having different types of singularities become
fully effective only when
solutions of the CHE admit
both the Leaver and  the Whittaker-Ince limits.
Counter-examples are provided by
Hylleraas \cite{hylleraas} and
Jaff\'e's \cite{jaffe} solutions  which admit
none of these limits, as we can see by using Leaver's form for
such solutions \cite{leaver1}.

Now  we introduce the Whittaker-Hill
and the Mathieu equations which are particular cases of both the
CHE  and the DCHE \cite{decarreau1}. A
trigonometric (hyperbolic) form of the Whittaker-Hill equation
(WHE) is \cite{arscott,ince}
\begin{eqnarray}\label{whe}
\frac{d^2W}{du^2}+\kappa^2\left[\vartheta-
\frac{1}{8}\xi^{2}
-(p+1)\xi\cos(2\kappa u)+
\frac{1}{8}\xi^{2}\cos(4\kappa u)\right]W=0, \ \ (\mbox{WHE}).
\end{eqnarray}
If $u$ is a real variable, this
equation represents the usual WHE when $\kappa=1$ and
the modified WHE when $\kappa=i$.
On the other hand, the Mathieu equation has the
form \cite{McLachlan}
\begin{eqnarray}\label{mathieu}
\frac{d^2w}{du^2}+\sigma^2\big[a-2k^2\cos(2\sigma u)\big]w=0,
\qquad(\mbox{Mathieu equation}),
\end{eqnarray}
where ÃÂ $\sigma=1$ or $\sigma=i$ for the Mathieu or
modified Mathieu equation, respectively.
Some details about solutions for the WHE and
Mathieu equation regarded as CHE or DCHE are
given in Ref. \cite{eu3}.
The Mathieu equation  is also a particular
case of equation (\ref{incedche}) as shown
in section 3. Incidentally, the original Whittaker-Ince
limit \cite{humbert,ince} is obtained when
$\xi\rightarrow 0$, $p\rightarrow \infty$
so that $p\xi=2k^2$, $\kappa=\sigma$ and $\vartheta=a$
in the WHE.ÃÂ  This gives the Mathieu equation.

On the other hand, the solutions for the Heun equations,
and in particular for CHE
and DCHE,  take one of the following forms  \cite{ronveaux}
\begin{eqnarray}\label{tipos-de-solucoes}
\displaystyle \sum_{n} a_{n}\ f_n(z)=\sum_{n=-\infty}^{\infty}a_{n}\ f_n(z),\qquad
\displaystyle \sum_{n=0}^{\infty} a_{n}\ f_n(z),\qquad
\displaystyle \sum_{n=0}^{N} a_{n}\ f_n(z),
\end{eqnarray}
where the series coefficients $a_{n}$
satisfy three-term or higher order
recurrence relations, $f_{n}(z)$
is a function of the independent variable $z$, $N$ is a non-negative
integer and the symbol $\sum_{n}$ holds for summation
running from negative to positive infinity. These are called,
respectively, two-sided infinite series,
one-sided infinite series and finite series. The finite series
are also known as quasi-polynomial solutions,
quasi-algebraic solutions or Heun polynomials.

Expansions in two-sided infinite series  are  necessary
to assure the series convergence  when
there is no free constant in the Heun equations.
Thus, all the parameters of the CHE and DCHE
which rule the time-dependence
of Klein-Gordon and Dirac test-fields in some Friedmannian
spacetimes \cite{birrell,eu1} are determined from
conditions imposed on the spatial part of
the wave functions \cite{scho1,scho2}. Similarly,
in ÃÂ the scattering problem of ions by a finite
dipole \cite{leaver1} or by polarisable targets
\cite{buhring,eu2} all the parameters of  the radial
Schr\"odinger equation are known.

When some parameters of the Heun equations  assume
special values, one-sided infinite series truncate
on the right  giving expansions in finite series. These
Heun polynomials may be useful to
get solutions for quasi-exactly solvable (QES) problems
\cite{turbiner,turbiner1,ushveridze1,ushveridze2}.
In effect,  a problem is QES if admits solutions given
by finite-series whose coefficients
necessarily satisfy three-term  or higher order
recurrence relations \cite{kalnins}; in contrast, a
problem is exactly solvable
if its solutions are given by
(generalised) hypergeometric functions. Actually,
there are QES potentials for which the Schr\"odinger
equation leads to the CHE and DCHE as well as to
the general, biconfluent and triconfluent Heun
equations (see Appendix A).

Excepting possibly the Heun polynomials,
in general the solutions for the Heun equations do
not converge for the entire  range of the
independent variable. Then, it is necessary
to consider two or more solutions converging over
different domains and having the appropriate behaviours
at the singular points. It is also  necessary to
take into account the transformation rules which
generate new  solutions from a known
solution (these rules result from substitutions of variables
which preserve the form of the Heun equations
but modify their parameters).

We will start with a set of
three solutions for the CHE, represented by
two-sided infinite series which have
coefficients that satisfy three-term recurrence
relations. These solutions admit both the Leaver and
the Whittaker-Ince limits and so we can generate
sets of solutions for all the equations discussed above.
Solutions obtained from this set by means
of transformation rules are given in sections 2.4 and
3.3.

More precisely, in Sec. 2 we take two Leaver's solutions
in series of regular and irregular confluent
hypergeometric functions for CHE and introduce another
expansion in series of irregular confluent hypergeometric
functions - see the solutions given in
Eqs. (\ref{che-primeiro set}) and the recurrence relations
(\ref{che-rec-1a}) and (\ref{che-rec-1b}).
The expansion in series of regular
functions converges for any $z$, whereas the two
expansions in series of irregular functions converge
for $|z|>|z_0|$.  From the properties of
the three-term recurrence relations and of the
hypergeometric functions,
we shall find conditions which permit to write
one solution as a linear combination of the
others in the region $|z|>|z_0|$. These conditions also
assure that the series
coefficients of the three solutions
are proportional to each other.

We advance that the convergence of the expansion in regular
functions over the entire complex
plane does not dispense with the expansions in series
of irregular functions. In fact, if the
expansion in regular functions is a linear
combination of the others,
its behaviour  at $z=\infty$ must be
given by a linear combination of the two behaviours
given in equation (\ref{thome1}), since each
expansion in irregular functions corresponds to
one of those behaviours. However,
if for instance $i\omega z$ is real, one of the exponentials
$\exp{(\pm i\omega z)}$ goes to infinity. In this case,
to get a solution finite when $z\rightarrow \infty$ we
have to use one of the expansions in irregular functions.

The solutions for the limits of the CHE,
given in the above diagram, are obtained by the same
procedure used in Ref. \cite{eu3}, where a set
having only two solutions in terms of one-sided series
of confluent hypergeometric functions was considered.
Thus, in Sec. 3 we  find
a set of three expansions in
series of Bessel functions for Whittaker-Ince
limit (\ref{incegswe}) of the CHE,
the three solutions possessing exactly the same
series coefficients.
In Sec. 4 we find that the solutions for
the DCHE are again given by series of confluent
hypergeometric, while the solutions for Whittaker-Ince
limit of the DCHE are given by series of Bessel functions
once more.

In summary, in Sec. 2 we deal with the CHE, in Sec. 3 with
the Whittaker-Ince limit (\ref{incegswe}) of the CHE,
and in Sec. 4 with the DCHE and its limit (\ref{incedche}).
Sec. 5 presents some conclusions and points to problems
which are ruled by the equations of the above diagram. Appendix
A shows some relations among Heun equations
and quasi-exactly solvable problems, while Appendix
B gives some properties of the confluent hypergeometric
functions.

%
%
\section*{2. The Confluent Heun Equation (CHE)}
In this section we review the two Leaver solutions
in series of regular and irregular confluent
hypergeometric functions for the confluent
Heun equation \cite{leaver1}  and  introduce the extra
expansion in series of irregular confluent hypergeometric
functions. The fact that the expansion in terms of regular
functions converges
for any $z$  (the expansions in terms of irregular
functions converge for $|z|>|z_0|$) distinguishes
the present solutions from the Leaver
expansions in series of Coulomb
wave functions \cite{leaver1} since the latter converge
only for $|z|>|z_{0}|$ (if the series is doubly infinite).

Firstly, in Sec. 2.1, we recall some features of three-term
recurrence relations for the series coefficients
and supply  properties of the confluent hypergeometric
functions whichÃÂ  enter  the solutions for the
CHE and DCHE. We also write down a Baber-Hass\'e
solution in power series since this
is important to obtain finite-series solutions
as well as to cover the cases in which the expansions
in series of regular confluent hypergeometric
functions are not valid.

In Sec. 2.2 we analyse the set constituted
by the expansions in series of
confluent hypergeometric functions.
This is called the fundamental set of solutions because, by means
of the transformation rules it originates
other sets of solutions for the CHE and, by
way of limiting processes, it affords sets of solutions
for the other equations given in the schema  of the
first section. We find the conditions under which
one solution of the fundamental set
is given as a linear combination
of the others. After that, we truncate the two-sided series
from below in order to get one-sided series solutions as well.
In Sec. 2.3, we discuss the convergence of new
expansion in series of irregular confluent hypergeometric
functions and, finally, in Sec. 2.4 we show how we can
generate new sets of solutions by using the transformation
rules for the CHE.

\subsection*{2.1. General remarks and the Baber-Hass\'e expansions}
If the coefficients of
one-sided series solutions are denoted by $b_{n}$,
then the three-term recurrence
relations have the form
\begin{eqnarray}\label{recurrence1}
\alpha_{0}b_{1}+\beta_{0}b_{0}=0,
\qquad \alpha_{n}b_{n+1}+\beta_{n}b_{n}+
\gamma_{n}b_{n-1}=0\quad (n\geq1)
\end{eqnarray}
where $\alpha_{n}$, $\beta_{n}$ and $\gamma_{n}$
depend on the parameters of the
differential equation. This system of
homogeneous linear equations
has nontrivial solutions for $b_{n}$ only if the
determinant of the respective infinite tridiagonal
matrix vanishes. This demands the presence of
some arbitrary parameter
in the differential equation.
Equivalently, these recurrence relations imply a
characteristic equation given by the infinite continued
fraction \cite{leaver1}
\begin{eqnarray}\label{charcteristic1}
\beta_{0}=\frac{\alpha_{0}\gamma_{1}}{\beta_{1}-}\
\frac{\alpha_{1}\gamma_{2}}
{\beta_{2}-}\ \frac{\alpha_{2}\gamma_{3}}{\beta_{3}-}\cdots,
\end{eqnarray}
which must be satisfied in order to assure the
series convergence.

If $\gamma_{n}=0$ for some $n=N+1$, where $N$ is a
non-negative integer, the one-sided series
terminates at $n=N$ and, consequently, gives a finite-series
solution with $0\leq n\leq N$ \cite{arscott}.
Thus the recurrence relations can be written in the form
\begin{eqnarray}
\label{matriz}
\left(
\begin{array}{ccccccccc}
\beta_{0} & \alpha_{0} &0 ÃÂ  &\cdotsÃÂ  & ÃÂ  & ÃÂ  ÃÂ  & ÃÂ & ÃÂ  & ÃÂ  0ÃÂ  ÃÂ  \\
\gamma_{1}&\beta_{1} ÃÂ  & \alpha_{1} & ÃÂ  ÃÂ  ÃÂ  & ÃÂ  ÃÂ  &
& ÃÂ & ÃÂ \\
ÃÂ  ÃÂ 0 ÃÂ  ÃÂ &\gamma_{2} & \beta_{2} ÃÂ  ÃÂ &\alpha_{2}& ÃÂ  &ÃÂ  & ÃÂ & ÃÂ ÃÂ  ÃÂ & ÃÂ  ÃÂ \\
ÃÂ \vdots ÃÂ & ÃÂ  ÃÂ  ÃÂ  ÃÂ  & ÃÂ  ÃÂ  ÃÂ  ÃÂ  ÃÂ  ÃÂ & ÃÂ   ÃÂ  ÃÂ  ÃÂ  ÃÂ & ÃÂ  &ÃÂ  & ÃÂ & ÃÂ  ÃÂ  ÃÂ  ÃÂ \\
& ÃÂ  & ÃÂ  & ÃÂ  & ÃÂ  & ÃÂ  & ÃÂ  & ÃÂ  & \\
ÃÂ  ÃÂ  ÃÂ ÃÂ  ÃÂ & ÃÂ ÃÂ & ÃÂ  & ÃÂ  & ÃÂ ÃÂ  & &ÃÂ \gamma_{N-1}& \beta_{N-1}&\alpha_{N-1}\\
ÃÂ ÃÂ 0 ÃÂ &\cdots ÃÂ  ÃÂ & ÃÂ  & ÃÂ  & ÃÂ  ÃÂ  ÃÂ  & ÃÂ  &ÃÂ  ÃÂ  0 ÃÂ  ÃÂ &\gamma_{N} & \beta_{N}
\end{array}
\right) \left(\begin{array}{l}
b_{0} ÃÂ \\
b_{1} \\
b_{2} \\
ÃÂ \vdots ÃÂ  \\
\\
\\
b_{N-1}\\
b_{N}
\end{array}
\right)=0.
\end{eqnarray}
If the elements $\alpha_{i}$,
$\beta_{i}$ and ÃÂ $\gamma_{i}$
of the previous matrix are real and if
\begin{eqnarray}\label{lemma}
\alpha_{i}\gamma_{i+1}> 0, \ \ ÃÂ \ \ 0\leq i\leq N-1,
\end{eqnarray}
then all the $N+1$ roots of its determinant are
real and different \cite{arscott}. This theorem
is important to determine a part of the energy
spectra in the case of quasi-exact potentials.
On the other hand, if $\alpha_{n}=0$ for
some $n=N$, the series begins at $n=N+1$,
but in this case one may set $n= m+N+1$
and rename the series coefficients in order
to obtain a series beginning at $m=0$.

Now, suppose that there is a second solution with coefficients
$c_{n}$ satisfying
\begin{eqnarray}\label{recurrence2}
\tilde{\alpha}_{0}c_{1}+\beta_{0}c_{0}=0,
\qquad \tilde{\alpha}_{n}c_{n+1}+\beta_{n}c_{n}+
\tilde{\gamma}_{n}c_{n-1}=0\quad (n\geq1),
\end{eqnarray}
where $\beta_{n}$ is the same as in Eq. (\ref{recurrence1}). Then,
if
\begin{eqnarray}\label{charcteristic2}
\tilde{\alpha}_{n}\tilde{\gamma}_{n+1}=\alpha_{n}\gamma_{n+1},
\end{eqnarray}
it follows from Eq. (\ref{charcteristic1})  that
both solutions have the same characteristic equation
if $n$ takes the same values in both series:
in these circumstances, $b_{n}$ and $c_{n}$ in
general are proportional to each other. Such
proportionality requires
the same range for $n$ in
Eqs. (\ref{recurrence1}) and (\ref{recurrence2}) because
there are cases in which the relation (\ref{charcteristic2})
is formally satisfied, but one solution is given by a
finite series while the other is given by an infinite series.
In these cases one series breaks off on the right and the
other on the left.

The previous remarks can be extended to
doubly infinite (or two-sided) series.
These expansions present a
parameter $\nu$ which must be determined
from a characteristic equation if there is no free
parameter in the differential equation, or can be chosen
at will if there is a free constant.
The recurrence relations for  the series
coefficients $b_{n}$ now take the form
\letra
\begin{eqnarray}\label{rec}
\alpha_{n}b_{n+1}+\beta_{n}b_{n}+\gamma_{n}b_{n-1}=0,
\quad (-\infty<n<\infty)
\end{eqnarray}
where $\alpha_{n}$, $\beta_{n}$, $\gamma_{n}$
and $b_{n}$ depend on the parameters
of the differential equation as well as on $\nu$.
These recurrence relations lead
to the characteristic equation \cite{leaver1}
\begin{eqnarray}\label{twosided-charact}
\beta_{0}=\displaystyle\frac{\alpha_{-1}\ \gamma_{0}}{\beta_{-1}-}
\ \frac{\alpha_{-2}\
\gamma_{-1}}{\beta_{-2}-}\
\frac{\alpha_{-3}\ \gamma_{-2}}
{\beta_{-3}-}\cdots+\frac{\alpha_{0}\ \gamma_{1}}{\beta_{1}-}
\ \frac{\alpha_{1}\ \gamma_{2}}
{\beta_{2}-}\
\frac{\alpha_{2}\ \gamma_{3}}{\beta_{3}-}\cdots.
\end{eqnarray}

If there is a second doubly infinite
series having the
recurrence relation $\tilde{\alpha}_{n}c_{n+1}+\beta_{n}c_{n}+
\tilde{\gamma}_{n}c_{n-1}=0$ such that the condition
(\ref{charcteristic2}) is fulfilled, then both solutions satisfy
the same characteristic equation (\ref{twosided-charact}).
The two series are really doubly infinite  if neither the
coefficients of $b_{n+1}$ and $c_{n+1}$ nor the
coefficients of  $b_{n-1}$ and $c_{n-1}$
vanish, since such conditions assure that the summation
extends from negative to positive infinity in both solutions.
This requires that
\antiletra
\begin{eqnarray}\label{firstcondition}
\alpha_{n}, \ \tilde{\alpha}_{n},
\  \gamma_{n}\text{ and } \tilde{\gamma}_{n} \text{ do not vanish for any } n,
\end{eqnarray}
a requirement which imposes constraints on the parameters of
differential equation and on the characteristic parameter
$\nu$, in the case of two-sided series. These conditions
will be useful for studying the sets of two-sided solutions of
the CHE and DCHE. In addition, if we choose $\nu$ such that
$\alpha_{-1}=\tilde{\alpha}_{-1}=0$, then the series
are truncated on the left since the summation
begins at $n=0$. Thus, we have
\begin{eqnarray}\label{truncamento}
\alpha_{-1}=\tilde{\alpha}_{-1}=0\quad
\Rightarrow\quad\text{one-sided series with }n\geq 0.
\end{eqnarray}

Other restrictions on $\nu$ and on the parameters of
the Heun equations come from the properties of the special functions
used to construct the series solutions. Thus, let us consider
expansions in series of regular and irregular confluent
hypergeometric functions for the CHE, denoted by $\Phi(a,c;y)$
and $\Psi(a,c;y)$ respectively. These are solutions of
the confluent hypergeometric equation \cite{erdelyi1}
\begin{eqnarray}\label{confluent0}
y\frac{d^2\varphi}{dy^2}+(c-y)\frac{d\varphi}{dy}-
a\ \varphi=0,
\end{eqnarray}
where the parameters $a$ and $c$ will depend on
summation index $n$, on the
parameters of the Heun equations and also on the characteristic
parameter $\nu$ in the case of two-sided  infinite series.
In fact, the following four solutions for Eq. (\ref{confluent0})
\begin{eqnarray}\label{confluent1}
\begin{array}{ll}
\varphi_{n}^{1}(y)=\Phi(a,c;y),
&\varphi_{n}^{2}(y)=\Psi(a,c;y),\vspace{1.5mm}\\
%
\varphi_{n}^{3}(y)=e^{y}\ y^{1-c}\Phi(1-a,2-c;-y),&
\varphi^{4}_{n}(y)=e^{y}\ y^{1-c}\ \Psi(1-a,2-c;-y)
\end{array}
\end{eqnarray}
are all of them defined and distinct only if
$c$ is not an integer \cite{erdelyi1}. Furthermore, if
\begin{eqnarray}\label{secondconditionB}
a,\ c \text{ and }c-a \text{ are not integer },
\end{eqnarray}
then any two of the solutions (\ref{confluent1})
form a fundamental system of solutions for
confluent hypergeometric equation \cite{erdelyi1}.
The formula
\begin{eqnarray}\label{continuation1}
\Psi(a,c;y)=\frac{\Gamma({1-c})}
{\Gamma({a-c+1})}\ \Phi(
a,c;y)+\frac{\Gamma(c-1)}
{\Gamma(a)}\ y^{1-c}\ \Phi(
a-c+1,2-c;y),
\end{eqnarray}
gives the analytic continuation of $\Psi$ in terms
of $\Phi$. The expression of $ \Phi$ in terms of $\Psi$ is
obtained from the previous one by using
the relation $\Gamma(z)\Gamma(1-z)=\pi/\sin(\pi{z})$.
One finds \cite{erdelyi1},
\begin{eqnarray}\label{continuation2}
\Phi(a,c;y)= \frac{e^{i\pi a\varepsilon}\Gamma(c)}
{\Gamma(c-a)}\  \Psi(a,c;y)+ \frac{e^{i\pi \varepsilon(a-c)}\Gamma(c)}
{\Gamma(a)}\
e^{y}\ \Psi(c-a,c;-y),\quad (\varepsilon=\pm 1)
\end{eqnarray}
where the plus or minus signs are to be taken throughout
following the conventions
\begin{eqnarray*}
\varepsilon=1 \ \text{if}\ (-1)=e^{i\pi},\ \
\varepsilon=-1 \ \text{if}\
(-1)=e^{-i\pi} \Rightarrow
(-1)^{c}=e^{i\pi{c}},\ \
(-1)^{c}=e^{-i\pi{c}}
\end{eqnarray*}
(by setting $c=1/2$ we see that $\varepsilon=1$ if
we take $\sqrt{-1}=i$,
and $\varepsilon=-1$ if $\sqrt{-1}=-i$).
The relation (\ref{continuation1}) allows writing $\Phi(a,c;y)$
as a combination of a regular and an irregular confluent hypergeometric
functions; analogously, Eq. (\ref{continuation2}) gives $\Psi(a,c;y)$
in terms of regular and irregular functions.


The general form of the
expansions in series of confluent hypergeometric
functions for the CHE and DCHE is
\begin{eqnarray}\label{fundamental-0}
U\big(\varphi_{n}^{i}(y)\big)=e^{-i\omega z}\displaystyle\sum_{n}
\pi_{n}^{i}\ \varphi_{n}^{i}(y),
\end{eqnarray}
where $\pi_{n}^{i}$ denotes the series coefficients.
In the fundamental set of solutions (\ref{che-primeiro set})
the argument of the hypergeometric functions
$\varphi_{n}^{i}$ is $y=2i\omega z$, and the
parameters are $a=({B_{2}}/{2})-i\eta$ and
$c=n+\nu+B_{2}$. We use only the functions
$\varphi_{n}^{1}(y)$, $\varphi_{n}^{2}(y)$
and $\varphi_{n}^{4}(y)$ because $\varphi_{n}^{3}(y)$
would lead to a solution whose domain of convergence
excludes all the singular points of the equation.
As $\varphi_{n}^{3}(y)$ is discarded,
the formula  (\ref{continuation2})
is the only one necessary to connect the
three hypergeometric functions.
Consequently, the conditions (\ref{secondconditionB})
are replaced by
\begin{eqnarray}\label{secondcondition}
a,\ c \text{ and }c-a \text{ are not zero
or {\bf negative} integers}.
\end{eqnarray}
The  present conditions combined with
conditions  (\ref{firstcondition}) allow to use
Eq. (\ref{continuation2}) in order to
write one solution of the CHE or DCHE
as a linear combination of the others.

Now we write down a
Baber-Hass\'e solution in series of ($z-z_0$) for the CHE
\cite{baber,eu3,leaver1} (which converges for finite
values of $z$), namely,
\letra
\begin{eqnarray}\label{baber}
U_{1}^{\text{baber}}(z)=e^{i\omega z}\displaystyle \sum_{n=0}^{\infty}a_{n}^{(1)}
(z-z_{0})^{n},\qquad (|z|=\text{finite})
\end{eqnarray}
where recurrence relations for the
coefficients are given by ($a_{-1}^{(1)}=0$)
\begin{eqnarray}\label{baber-recorrencia}
&&ÃÂ z_{0}\left(n+B_{2}+\frac{B_{1}}{z_{0}}\right)
\left(n+1\right)a_{n+1}^{(1)}+
\bigg[ n\left(n+B_{2}-1+2i\omega z_{0}\right)\nonumber\\
%
%
&&+B_{3}+ i\omega z_{0}\left(B_{2}+
\frac{B_{1}}{z_{0}}\right)\bigg]a_{n}^{(1)}+
2i\omega\left(n+i\eta+\frac{B_{2}}{2}-1\right)
a_{n-1}^{(1)}=0.\hspace{1.5cm}
\end{eqnarray}
This expansion provides a 
solution for the case in which
the Leaver expansion in regular hypergeometric functions,
given in the following,
is not valid, and admits both the Leaver
and the Whittaker-Ince limits.
Furthermore, it gives  finite-series solutions
with $0\leq n\leq N$  when $i\eta+B_2/2=-N$.
However, the Whittaker-Ince limit of this Baber-Hass\'e
solution does not admit finite-series solutions
since the coefficient of $a_{n-1}$ is a constant:
$2i\omega\left(n+i\eta+B_2/2-1\right)\rightarrow
q$. 
ÃÂ 
%
%
%
%
%
We may generate a group
containing 16 sets of solutions for the CHE by applying to the
previous solution the transformation rules
given in Sec. 2.4.
These solutions for the CHE give  solutions
for the Whittaker-Hill equation (WHE).
In effect, if  $U(z)$ symbolises the solutions for the CHE, the solutions
$W(u)$ for the WHE  (\ref{whe})
are obtained by writing \cite{eu3}
\antiletra\letra
\begin{eqnarray}\label{wheasgswe}
W(u)=U(z), \qquad z=\cos^{2}(\kappa u),\qquad ÃÂ  (\kappa=1,i)
\end{eqnarray}
where the parameters of $U(z)$ are given in
terms of the parameters of the WHE by
\begin{eqnarray}
z_{0}=1,\ \ B_{1}=-\frac{1}{2}, \quad B_{2}=1, \ \
B_{3}=\frac{1}{4}[(p+1)\xi-\vartheta], \quad i\omega=\frac{\xi}{2},
\ \ i\eta=\frac{p+1}{2}.
\end{eqnarray}
For the WHE, the
Baber-Hass\'e solutions become
even or odd solutions
with respect to the change of the sign of $u$
\cite{eu3}.

\subsection*{2.2. The fundamental set of solutions}
First we examine the two-sided infinite series solutions
and  then we obtain the one-sided series.
It is worth advancing that the solutions in terms of regular
confluent hypergeometric functions $\Phi(a,c;y)$
are not valid when
the first parameter $a$ is zero or a negative integer.
This statement is true only if the solutions are written
so that the parameter $a$
does not depend on the summation index $n$, since
the Kummer relation (\ref{kummer}) gives another
representation in which the first parameter
of $\Phi(a,c;y)$ depends on $n$.

Defining the function $\tilde{\Phi}(a,b;y)$ by \cite{leaver1}
\antiletra
\begin{eqnarray}\label{phi-tilde}
\tilde{\Phi}(a,c;y)=\frac{\Gamma(c-a)}{\Gamma(c)}\ \Phi(a,c;y)
=\frac{\Gamma(c-a)}{\Gamma(c)}\left[1+\frac{a}{1!c}y+
\frac{a(a+1)}{2!c(c+1)}y^2+\cdots\right],
\end{eqnarray}
the fundamental set (\ref{fundamental-0}) reads
\letra
\begin{eqnarray}\label{che-primeiro set}
\begin{array}{l}
U_{{1}}(z)=e^{-i\omega z}\displaystyle\sum_{n}
(-1)^{n}b_{n}^{(1)}
\ \tilde{\Phi}\left(\frac{B_{2}}{2}-i\eta,
n+\nu+B_{2};2i\omega z\right),\vspace{1.5mm}\\
%
 U_{{1}}^{\infty}(z)=e^{-i\omega z}\displaystyle
\sum_{n}
(-1)^n b_{n}^{(1)}
\Psi\left(\frac{B_{2}}{2}-i\eta,n+\nu+B_{2};
2i\omega z\right),\vspace{1.5mm}\\
%
\bar U_{{1}}^{\infty}(z)=e^{i\omega z}\times\vspace{1.5mm}\\ \hspace{1.5cm}
%
\displaystyle\sum_{n}
c_{n}^{(1)}(-2i\omega{z})^{1-n-\nu-B_2}\ \Psi\left(1+i\eta-\frac{B_{2}}{2},
2-n-\nu-B_{2};-2i\omega z\right)
\end{array}
\end{eqnarray}
where the recurrence relations for $b_{n}^{(1)}$
and $c_{n}^{(1)}$ are
\begin{eqnarray}\label{che-rec-1a}
&& (n+\nu+1)\left(n+\nu+i\eta+\frac{B_2}{2}\right)b_{n+1}^{(1)}+
\bigg[(n+\nu)(n+\nu+B_{2}-1+2i\omega z_{0})\nonumber\\
%
&& +B_3+i\omega
z_0 \left(B_{2}+\frac{B_{1}}{z_{0}}\right)\bigg]
b_{n}^{(1)}+2i\omega
z_0\left(n+\nu+B_2+\frac{B_1}{z_0}-1\right)b_{n-1}^{(1)}=0.
\hspace{1cm}
\end{eqnarray}
and
\begin{eqnarray}\label{che-rec-1b}
(n+\nu+1)c_{n+1}^{(1)}+
\bigg[(n+\nu)(n+\nu+B_{2}-1+2i\omega z_{0})+i\omega
z_{0}\left(B_{2}+\frac{B_{1}}{z_{0}}\right)\hspace{.7cm} \nonumber\\
%
+B_{3}
\bigg]c_{n}^{(1)}+2i\omega
z_{0}\left(n+\nu+B_{2}+\frac{B_{1}}{z_{0}}
-1\right)\left(n+\nu+i\eta+\frac{B_{2}}{2}-1\right)
c_{n-1}^{(1)}=0.\
\end{eqnarray}
Apart from a multiplicative constant, the
coefficients $b_{n}^{(1)}$ and $c_{n}^{(1)}$
are connected by
\antiletra
\begin{eqnarray}\label{proporcional}
&c_{n}^{(1)}=
\Gamma\left(n+\nu+i\eta+\frac{B_2}{2} \right)b_{n}^{(1)},
\end{eqnarray}
provided that the argument of the gamma function is
not zero or negative integer.

The solutions $U_{1}(z)$ and $U_{1}^{\infty}(z)$
have been taken from Eqs. (166) and (167) of
Leaver's paper \cite{leaver1}
with $\nu$ replaced by $\nu+B_{2}$. On the other hand,
by using a Kummer relation given in
Eq. (\ref{kummer}), $\bar{U}_{1}^{\infty}(z)$  is
rewritten as
\begin{eqnarray}
\bar U_{{1}}^{\infty}(z)=e^{i\omega z}
\displaystyle\sum_{n}
c_{n}^{(1)} \Psi\left(n+\nu+i\eta+\frac{B_{2}}{2},
n+\nu+B_{2};-2i\omega z\right).
\end{eqnarray}
Then, it becomes clear that this solution can be
obtained by
substituting $n+\nu$ for
$n$ in the one-sided solution given in Eq. (33a) of
Ref. \cite{eu3} and by allowing that the
summation runs from negative to positive infinity.
The solution $U_{1}$ converges
for any $z$ \cite{leaver1}, while
both $U_{1}^{\infty}$
and $\bar{U}_{1}^{\infty}$
converge for $|z|>|z_{0}|$. From
the fact that $\Phi(a,c;0)=1$ and
from Eq. (\ref{asymptotic-confluent1}), it follows that
\begin{eqnarray}\label{par1-infinito}
\lim_{z\rightarrow 0}U_{1}(z)\sim 1,\quad
\lim_{z\rightarrow \infty}U_{1}^{\infty}(z)\sim e^{-i\omega z}
z^{i\eta-\frac{B_2}{2}},
\quad\lim_{z\rightarrow \infty}\bar{U}_{1}^{\infty}(z)\sim e^{i\omega z}
z^{-i\eta-\nu-\frac{B_2}{2}}.
\end{eqnarray}
Thus, two different behaviours at $z=\infty$ are included
in the solutions belonging to the same set. Notice
that the parameter $\nu$ does not appear in Eq. (\ref{thome1})
because this was obtained from solutions
given by one-sided series (Thom\'e solutions).

Since we are dealing with three solutions for a second order
linear differential equation, now we establish the conditions
to get one of these as a linear combination of the others in
a domain where the three solutions are valid. From the recurrence
relations (\ref{che-rec-1a})
and (\ref{che-rec-1b}) we find that the three series are
really doubly infinite if
\letra
\begin{eqnarray}\label{serieinfinita-1}
&\nu,
\ \nu+B_{2}+\frac{B_{1}}{z_{0}}\text{ and }
\nu+i\eta+\frac{B_{2}}{2} \text{ are not integers, }
\end{eqnarray}
since under these conditions neither the coefficients
of $b_{n}^{(1)}$
and $c_{n}^{(1)}$ nor the coefficients of $b_{n-1}^{(1)}$ and $c_{n-1}^{(1)}$
vanish, that is to say,
the series do not truncate on the left or on the right.
The last condition also assures that the series
coefficients are linked by Eq. (\ref{proporcional})
and in turn this implies that there is a unique characteristic
equation. If, in addition to conditions (\ref{serieinfinita-1}),
the conditions (\ref{secondcondition}) are also satisfied,
that is, if
\begin{eqnarray*}
&n+\nu+B_{2},\
\frac{B_{2}}{2}-i\eta \text{ and }
n+ \nu+i\eta+\frac{B_{2}}{2}\ \text{are not
zero or negative integers,}
\end{eqnarray*}
then Eq. (\ref{continuation2}) may be used to prove
that any of the three solutions is a linear combination
of the others in a region where the three solutions are
valid ($|z|>|z_0|$). Since $n$ runs from $-\infty$  to $\infty$,
the above conditions are equivalent to
\begin{equation}\label{serieinfinita-2}
\nu+B_{2}, \  \nu+i\eta+\frac{B_{2}}{2} \text{ are not integers}; \
\frac{B_{2}}{2}-i\eta \text{ is not
zero or negative integer},
\end{equation}
which repeat one of the conditions (\ref{serieinfinita-1}).

In short, to express one solution in terms of the others,
the three solutions must be given by
two-sided series and the formula (\ref{continuation2}) for
analytic continuation of the hypergeometric functions
must hold. These are the general conditions which
may be applied to any set of solutions generated from
the first set through the transformation rules of the CHE.
In fact, they are equivalent to the conditions
(\ref{firstcondition}) and (\ref{secondcondition}).

If $U_{1}$ is a superposition
of $U_{1}^{\infty}$ and $\bar{U}_{1}^{\infty}$
in the common domain of convergence ($|z|>|z_0$),
then the behaviour of $U_{1}$ when $z\rightarrow\infty$
must be given by a combination of the behaviours
of $U_{1}^{\infty}$ and $\bar{U}_{1}^{\infty}$.
However, for certain problems, one of the
expansions in irregular functions may be inadequate
when $z\rightarrow\infty$ and, consequently,
$U_{1}$ becomes inappropriate as well.

Only the restriction on the values of $(B_{2}/2)-i\eta$ cannot
be satisfied by a convenient choice of $\nu$. This restriction also
arises if we  consider the solution $U_1(z)$ by itself,
disregarding its connection with the other solutions.  In fact, if
$(B_{2}/2)-i\eta=-m$  ($m=0,1,2,\cdots$), the hypergeometric
function $\Phi(a,c;y)$ which appears in $U_{1}$ becomes
a polynomial of degree $m$ with respect to
its argument \cite{abramowitz} and,
then, the summation from negative to positive infinity
is meaningless. The solution
\antiletra \letra
\begin{eqnarray}\label{baber-p}
U_{1}^{\text{p}}(z)=e^{-i\omega z}\displaystyle \sum_{n=0}^{\infty}d_{n}^{(1)}
(z-z_{0})^{n},\qquad (|z|=\text{finite})
\end{eqnarray}
where recurrence relations for the
coefficients are ($d_{-1}^{(1)}=0$)
\begin{eqnarray}\label{baber-recorrencia-p}
&&z_{0}\left(n+B_{2}+\frac{B_{1}}{z_{0}}\right)
\big(n+1\big)d_{n+1}^{(1)}+
\bigg[ n\left(n+B_{2}-1-2i\omega z_{0}\right)\nonumber\\
%
&&+B_{3}- i\omega z_{0}\left(B_{2}+
\frac{B_{1}}{z_{0}}\right)\bigg]d_{n}^{(1)}
-2i\omega\left(n-i\eta+\frac{B_{2}}{2}-1\right)
d_{n-1}^{(1)}=0,\hspace{1cm}
\end{eqnarray}
takes the place of $U_{1}$ when $(B_{2}/2)-i\eta=-m$.
Notice that  $U_{1}^{\text{p}}$ was obtained from the
Baber-Hass\'e expansion (\ref{baber}) by substituting
 ($-\omega,-\eta$) for ($\omega,\eta$). Furthermore,
even if $(B_{2}/2)-i\eta=-m$,
by means of the transformation rules we can find
two-sided series expansions in terms of regular
confluent hypergeometric functions which
hold for this case.

%
Now we consider the one-sided series solutions. From the
recurrence relations (\ref{che-rec-1a})
and (\ref{che-rec-1b}) we see that for
truncating  the three solutions on the left at $n=0$
the only choice of $\nu$ common
to the three solutions  is $\nu=0$ -- see Eq. (\ref{truncamento}).
We rewrite these one-sided series solutions as
\antiletra\letra
\begin{eqnarray}\label{truncada1}\begin{array}{l}
U_{1}(z)=e^{-i\omega z}\displaystyle\sum_{n=0}^{\infty}
\frac{(-1)^{n}c_{n}^{(1)}}{\Gamma(n+B_2)}
\ {\Phi}\left(\frac{B_{2}}{2}-i\eta,n+B_{2};2i\omega z\right),\vspace{1.5mm}\\
%
%
U_{1}^{\infty}(z)=e^{-i\omega z}\displaystyle\sum_{n=0}^{\infty}
b_{n}^{(1)}
\ \Psi\left(\frac{B_2}{2}-i\eta,n+B_{2};2i\omega z\right),\vspace{1.5mm}\\
%
\bar{U}_{1}^{\infty}(z)=e^{i\omega z}\displaystyle\sum_{n=0}^{\infty}
c_{n}^{(1)}
\ {\Psi}\left(n+i\eta+\frac{B_2}{2},n+B_{2};-2i\omega z\right),
\end{array}
\end{eqnarray}
where the recurrence relations are
\big($b_{-1}^{(1)}=c_{-1}^{(1)}=0$\big):
\begin{eqnarray}\label{truncadaR1}
(n+1)\left(n+i\eta+\frac{B_2}{2}\right)b_{n+1}^{(1)}+
\beta_n^{(1)} b_{n}^{(1)}+
2i\omega
z_0\left(n+B_2+\frac{B_1}{z_0}-1\right)b_{n-1}^{(1)}=0,
\end{eqnarray}
and
\begin{eqnarray}\label{truncadaR2}
&& (n+1)c_{n+1}^{(1)}+\beta_{n}^{(1)}
c_{n}^{(1)}\nonumber\\
%
&& +2i\omega
z_{0}\left(n+B_{2}+\frac{B_{1}}{z_{0}}
-1\right)\left(n+i\eta+\frac{B_{2}}{2}-1\right)
c_{n-1}^{(1)}=0,
\end{eqnarray}
where
\begin{eqnarray*}
&\beta_{n}^{(1)}=n(n+B_{2}-1+2i\omega z_{0})+B_{3}+i\omega
z_{0}\left( B_{2}+\frac{B_{1}}{z_{0}}\right) .
\end{eqnarray*}
According to the previous subsection, if
$i\eta+(B_{2}/2)=-l$ ($l=0,1,2,\cdots$),
the series in ${U}_{1}$ and
$\bar{U}_{1}^{\infty}$ break off on the
right and these solutions reduce to Heun
polynomials ($0\leq{n}\leq{l}$),
while the solution $U_{1}^{\infty}$ truncates
on the left ($n\geq{l+1}$).
However, there is no need of considering these
Heun polynomials since the Baber-Hass\'e
solution (\ref{baber}) also suplies finite-series
solutions with the same characteristic equation.

For the solutions (\ref{truncada1}) the series
are infinite, with $0\leq n<\infty$,
if
\antiletra\letra
\begin{eqnarray}
&B_{2}+\frac{B_{1}}{z_{0}}\text{ and }
i\eta+\frac{B_{2}}{2} \text{ are not zero
or negative integers,}
\end{eqnarray}
as we see from the recurrence relations (\ref{truncadaR1}) and
(\ref{truncadaR2}). The solutions can be connected by means of
Eq. (\ref{continuation2}) if
\begin{eqnarray}
&B_{2},\
\frac{B_{2}}{2}-i\eta\text{ and }
i\eta+\frac{B_{2}}{2}\text{ are not
zero or negative integers,}
\end{eqnarray}
as we see from conditions (\ref{secondcondition}).
Then, the relation $c_{n}^{(1)}=
\Gamma[n+i\eta+({B_2}/{2}) ]b_{n}^{(1)}$
is well defined  and Heun polynomials are excluded from
(\ref{truncada1}). Under the above conditions,
the coefficients of the Baber-Hass\'e solution (\ref{baber})
are proportional to $c_{n}^{(1)}$, that is,
\antiletra
\begin{eqnarray}
a_{n}^{(1)}=\frac{c_{n}^{(1)}}{(z_{0})^{n}
\Gamma[n+B_2+(B_{1}/z_{0})]}.
\end{eqnarray}
This gives another reason to consider the
Baber-Hass\'e solutions in conjunction with
the expansions in confluent hypergeometric functions.

Finally, notice that there is another type of finite series
which is not included in the Baber-Hass\'e
expansions. It occurs when
\begin{eqnarray}
&B_{2}+\frac{B_{1}}{z_{0}}=-l,\qquad l=1,2,3,\cdots,
\end{eqnarray}
in which case the three series in (\ref{truncada1})
terminate at $n=l$. These Heun polynomials
are important because they are preserved by the
Whittaker-Ince limit, in
opposition to the Heun polynomials obtained
from the Baber-Hass\'e expansions.

%
%

\subsection*{2.3. Convergence of the third solution}
For one-sided infinite series the convergence of
$\bar{U}_{1}^{\infty}$ have already
been established in Ref. \cite{eu3}. Next
we show that the two-sided infinite series  converges
in both directions, that is,
when $n\rightarrow \infty$ and when $n\rightarrow -\infty$.
First we write the solution as
\begin{eqnarray}\label{psi-n}
\bar{U}_{1}^{(\infty)}(z)=
e^{i\omega z}\displaystyle \sum_{n=-\infty}^{\infty}
c_{n}^{(1)}y^{1-B_{2}-n-\nu}\Psi_{n}(y),
\end{eqnarray}
where%
\begin{eqnarray*}
\Psi_{n}(y)=
\Psi\left(1+i\eta-\frac{B_{2}}{2},
2-n-\nu-B_{2};y\right), \quad y=-2i\omega z.
\end{eqnarray*}
To determine the convergence
of the series, we have to find the ratios
\begin{eqnarray}\label{ratio}
\lim_{n\rightarrow \infty}\frac{c_{n+1}^{(1)}\ \Psi_{n+1}(y)}
{c_{n}^{(1)}\ y\ \Psi_{n}(y)},
\qquad\lim_{n\rightarrow -\infty}\frac{c_{n-1}^{(1)}\ y\  \Psi_{n-1}(y)}
{c_{n}^{(1)}\ \Psi_{n}(y)}.
\end{eqnarray}
For this, in the first place we divide the recurrence
relations (\ref{che-rec-1b})  by
$n c_{n}^{(1)}$ and retain only the leading  terms,
that is,
\begin{eqnarray*}
&& \left[1+\frac{\nu+1}{n}\right]
\frac{c_{n+1}^{(1)}}{c_{n}^{(1)}}+\left[
n+2\nu+B_{2}-1+2i\omega z_{0}+O\left(\frac{1}{n} \right)\right]\\
%
&&+\displaystyle 2i\omega
z_{0}\left[ n+2\nu+\frac{3}{2}B_{2}+\frac{B_{1}}{z_{0}}
+i\eta-2+O\left(\frac{1}{n} \right) \right]
\frac{c_{n-1}^{(1)}}{c_{n}^{(1)}}=0.
\end{eqnarray*}
The minimal solutions for this equation are
\begin{eqnarray}\begin{array}{l}
\displaystyle\lim_{n\rightarrow+\infty}\frac{c_{n+1}^{(1)}}{c_{n}^{(1)}}=
-2i\omega z_{0}
\left[1+\frac{1}{n}\left(i\eta-1+\frac{B_1}{z_0} +\frac{B_2}{2}
\right)  \right],\vspace{1.5mm}\\
%
%
\displaystyle\lim_{n\rightarrow-\infty}\frac{c_{n-1}^{(1)}}{c_{n}^{(1)}}=
-\frac{1}{n}
\left[1-\frac{B_2+\nu-3}{n}\right]. 
\end{array}
\end{eqnarray}
On the other hand, from Eqs. (\ref{relation2})
we get
\begin{eqnarray*}
-\left(n+\nu+i\eta+\frac{B_{2}}{2} \right) \Psi_{n+1}(y)+\big(n+\nu+B_{2}-1-y\big)
\Psi_{n}(y)+y\Psi_{n-1}(y)=0.
\end{eqnarray*}
Dividing this by $n\Psi_n$ we obtain
\begin{eqnarray*}
-\left[1+\frac{1}{n}\left( \nu+i\eta+\frac{B_{2}}{2}\right)
\right]\frac{\Psi_{n+1}}{\Psi_{n}} +
\left[ 1+\frac{\nu+
B_{2}-1-y}{n}\right] +\frac{y}{n}\frac{\Psi_{n-1}}{\Psi_{n}}=0.
\end{eqnarray*}
Then we can verify that
%
\begin{eqnarray}
\lim_{n\rightarrow+\infty}\frac{\Psi_{n+1}}{\Psi_{n}}=
1-\frac{1}{n}\left(1+i\eta-\frac{B_2}{2} \right)
 ,\quad\lim_{n\rightarrow-\infty}\frac{\Psi_{n-1}}{\Psi_{n}}=
-\frac{1}{y}
\left(n+\nu+B_2-1\right).
\end{eqnarray}
In fact there are other possibilities, but the preceding are
the only ones compatible with Eqs. (B.5)
and (B.6), respectively.
Hence, since $y=-2i\omega z$, we find
\begin{eqnarray}\begin{array}{l}
\displaystyle\lim_{n\rightarrow \infty}\frac{c_{n+1}^{(1)}\ \Psi_{n+1}(y)}
{c_{n}^{(1)}\ y\ \Psi_{n}(y)}=\frac{z_0}{z}\left[1+
\frac{1}{n}\left(B_2+\frac{B_1}{z_0} -2\right)  \right] ,\vspace{1.5mm}\\
%
\displaystyle\lim_{n\rightarrow -\infty}\frac{c_{n-1}^{(1)}\ y\  \Psi_{n-1}(y)}
{c_{n}^{(1)}\ \Psi_{n}(y)}=1+\frac{2}{n}.
\end{array}
\end{eqnarray}
Therefore, by the ratio test \cite{watson}
the series in Eq. (\ref{psi-n})
converges in the region $|z|>|z_0$.

%
\subsection*{2.4. Other sets of solutions for the CHE}
In order to generate other solutions for the CHE we
use the transformations rules resulting from
substitutions of variables which leave the form of the CHE
unaltered but change its parameters. Thus,
if $U(z)=U(B_{1},B_{2},B_{3}; z_{0},\omega,\eta;z)$
denotes one solution of the CHE in the Leaver form
(\ref{dche}), we have the  rules $T_{1},\ T_{2},\ T_{3}$ and
$T_{4}$  which operate as \cite{eu3}
%
\letra
\begin{eqnarray}\label{transformacao4}\begin{array}{ll}
T_{1}U(z)=z^{1+B_{1}/z_{0}}
U(C_{1},C_{2},C_{3};z_{0},\omega,\eta;z),& z_{0}\neq0,\vspace{1.5mm}\\
%
T_{2}U(z)=(z-z_{0})^{1-B_{2}-B_{1}/z_{0}}U(B_{1},D_{2},D_{3};
z_{0},\omega,\eta;z), & ÃÂ z_{0}\neq0,\vspace{1.5mm}\\
%
T_{3}U(z)=U(B_{1},B_{2},B_{3}; z_{0},-\omega,-\eta;z),
& \forall z_{0},\vspace{1.5mm}\\
%
T_{4}U(z)=
U(-B_{1}-B_{2}z_{0},B_{2},
B_{3}+2\eta\omega z_{0};z_{0},-\omega,
\eta;z_{0}-z),& ÃÂ \forall z_{0},
\end{array}
\end{eqnarray}
where
\begin{eqnarray}\label{constantes-C-D}
\begin{array}{l}
C_{1}=-B_{1}-2z_{0}, \ \
C_{2}=2+B_{2}+\frac{2B_{1}}{z_{0}},\ C_{3}=B_{3}+
\left(1+\frac{B_{1}}{z_{0}}\right)
\left(B_{2}+\frac{B_{1}}{z_{0}}\right),\vspace{1.5mm}\\
%
D_{2}=2-B_{2}-\frac{2B_{1}}{z_{0}},\ D_{3}=B_{3}+
\frac{B_{1}}{z_{0}}\left(\frac{B_{1}}{z_{0}}
+B_{2}-1\right).
\end{array}
\end{eqnarray}
Applying these rules to the basic set (\ref{che-primeiro set})
we may generate a group containing 16 sets of solutions
for the CHE. We will write only the subgroup obtained
by using the rules $T_{1}$ and $T_{2}$ in this order:
\antiletra
\begin{equation}\label{rules}
\left(U_{{1}},U_{{1}}^\infty,
\bar{U}_{{1}}^\infty\right)
\stackrel{T_1}{\longleftrightarrow}
\left(U_{{2}},U_{{2}}^\infty,
\bar{U}_{{2}}^\infty\right)
\stackrel{T_2}{\longleftrightarrow}
\left(U_{{3}},U_{{3}}^\infty,
\bar{U}_{{3}}^\infty\right)
\stackrel{T_1}{\longleftrightarrow}
\left(U_{{4}},U_{{4}}^\infty,
\bar{U}_{{4}}^\infty\right)
\end{equation}
where $\left(U_{{1}},U_{{1}}^\infty,
\bar{U}_{{1}}^\infty\right)$ denotes the first set of solutions
(\ref{che-primeiro set}).

In each of the following sets, the conditions for doubly
infinite series are obtained by
choosing $\nu$ such that the coefficients of $c_{n}^{(i)}$ and
$c_{n-1}^{(i)}$ do not vanish in the recurrence relations.
Besides this, one solution can be expressed as a
linear combination of the
others, if $a$, $c$ and $c-a$ are not negative integers,
where $a$ and $c$
are respectively the first and the second parameters of
the confluent hypergeometric functions: these conditions
lead to some restrictions on the values of $\nu$ as well as
on the parameters of the CHE. Despite the notation, it
is understood that the parameter $\nu$ may be different
in each set of solutions.

The second set of solutions admits the Whittaker-Ince limit
but does not admit the Leaver limit. It reads
\letra
\begin{eqnarray}\begin{array}{l}
U_{2}(z)=e^{-i\omega z}z^{1+\frac{B_{1}}{z_{0}}}
\times\vspace{1.5mm}\\
%
\hspace{1cm}
\displaystyle\sum_{n}
(-1)^{n}b_{n}^{(2)}
\widetilde{\Phi}\left(1-i\eta+\frac{B_{1}}{z_{0}}+\frac{B_{2}}{2},
n+\nu+2+B_{2}+\frac{2B_{1}}{z_{0}};2i\omega z\right),\vspace{1.5mm}\\
%
U_{2}^{\infty}(z)=e^{-i\omega z}z^{1+\frac{B_{1}}{z_{0}}}
\times\vspace{1.5mm}\\
%
\hspace{1cm}\displaystyle
\sum_{n}
(-1)^n b_{n}^{(2)}
\Psi\left(1-i\eta+\frac{B_{1}}{z_{0}}+\frac{B_{2}}{2},
n+\nu+2+B_{2}+\frac{2B_{1}}{z_{0}};2i\omega z\right),\vspace{1.5mm}\\
%
\bar{U}_{2}^{\infty}(z)=
e^{i\omega z}z^{1+\frac{B_{1}}{z_{0}}}\times\vspace{1.5mm}\\
%
\hspace{1cm}
\displaystyle\sum_{n}
c_{n}^{(2)}\Psi\left(n+\nu+1+i\eta+\frac{B_{1}}{z_{0}}+
\frac{B_{2}}{2},
n+\nu+2+B_{2}+\frac{2B_{1}}{z_{0}};-2i\omega z\right),
\end{array}
\end{eqnarray}
where the recurrence relations for $c_{n}^{(2)}$ are
\begin{eqnarray}
&&(n+\nu+1)c_{n+1}^{(2)}
+\bigg[(n+\nu)\left(n+\nu+1+2i\omega z_{0}+B_{2}
+\frac{2B_{1}}{z_{0}}\right)\nonumber\\
%
&&+B_{3}+\left(1+\frac{B_{1}}{z_{0}}\right)\left(B_{2}+
\frac{B_{1}}{z_{0}}\right)
+i\omega z_{0}\left(B_{2}+\frac{B_{1}}{z_{0}}\right)\bigg]c_{n}^{(2)}\nonumber\\
%
&&+2i\omega z_{0}\left(n+\nu+B_{2}+\frac{B_{1}}{z_{0}}-1\right)
\left(n+\nu+i\eta+\frac{B_{1}}{z_{0}}+\frac{B_{2}}{2}\right)
c_{n-1}^{(2)}=0.\hspace{1cm}
\end{eqnarray}
%
%
%
%
%
%
%
%
%
The recurrence relations for $b_{n}^{(2)}$ are
obtained from the previous ones by
\begin{eqnarray}
&c_{n}^{(2)}=
\Gamma\big[n+\nu+1+i\eta+\frac{B_{1}}{z_{0}}
+\frac{B_{2}}{2}\big] \ b_{n}^{(2)}.
\end{eqnarray}
The third set of solutions, which admits both limits, is
\antiletra\letra
\begin{eqnarray}\label{B3-a}
\begin{array}{l}
U_{3}(z)=e^{-i\omega z}(z-z_0)^{1-B_2-\frac{B_1}{z_0}}
\  z^{1+\frac{B_1}{z_0}}\times\vspace{1.5mm}\\
%
\hspace{1.4cm}\displaystyle\sum_{n}
(-1)^{n}b_{n}^{(3)}
\widetilde{\Phi}\left(2-i\eta-\frac{B_2}{2},
n+\nu+4-B_2;2i\omega z\right),\vspace{1.5mm}\\
U_{3}^{\infty}(z)=e^{-i\omega z}(z-z_0)^{1-B_2-\frac{B_1}{z_0}}
\ z^{1+\frac{B_1}{z_0}}\times\vspace{1.5mm}\\
%
\hspace{1.4cm}\displaystyle\sum_{n}
(-1)^n b_{n}^{(3)}\Psi\left(2-i\eta-\frac{B_2}{2},
n+\nu+4-B_2;2i\omega z\right),\vspace{1.5mm}\\
%
\bar{U}_{3}^{\infty}(z)=e^{i\omega z}(z-z_0)^{1-B_2-\frac{B_1}{z_0}}
\ z^{1+\frac{B_1}{z_0}}
\times \vspace{1.5mm}\\
%
%
\hspace{1.4cm}\displaystyle\sum_{n}
c_{n}^{(3)}\Psi\left(n+\nu+2+i\eta-\frac{B_{2}}{2},
n+\nu+4-B_{2};-2i\omega z\right),
\end{array}
\end{eqnarray}
where the $c_{n}^{(3)}$ satisfy 
\begin{eqnarray}\label{B3-b}
&&(n+\nu+1)c_{n+1}^{(3)}\nonumber\\
%
&&+\Big[(n+\nu)\left(n+\nu+3-B_2+2i\omega z_{0}\right)+
i\omega z_{0}\left(2-B_2-\frac{B_{1}}{z_{0}}\right)+
B_{3}+2-B_2
\Big]c_{n}^{(3)}\nonumber\\
%
&&+2i\omega z_0 \left(n+\nu+1-B_2-\frac{B_1}{z_0}\right)
\left(n+\nu+1+i\eta-\frac{B_2}{2}\right)c_{n-1}^{(3)}=0
\end{eqnarray}
and the recurrence relations for $b_{n}^{(3)}$
follow from
\begin{eqnarray}
&c_{n}^{(3)}=\Gamma\big[n+\nu+i\eta+2-\frac{B_2}{2}\big]\
b_{n}^{(3)}.
\end{eqnarray}

The fourth set of solutions is
\antiletra\letra
\begin{eqnarray}\begin{array}{l}
U_{4}(z)=e^{-i\omega z}(z-z_0)^{1-B_2-\frac{B_1}{z_0}}
\times\vspace{1.5mm}\\
%
\qquad\displaystyle\sum_{n}
(-1)^{n}b_{n}^{(4)}
\widetilde{\Phi}\left(1-i\eta-\frac{B_1}{z_0}-\frac{B_2}{2},
n+\nu+2-B_2-\frac{2B_1}{z_0};2i\omega z\right),\vspace{1.5mm}\\
%
U_{4}^{\infty}(z)=e^{-i\omega z}(z-z_0)^{1-B_2-\frac{B_1}{z_0}}
\times\vspace{1.5mm}\\
%
\qquad\displaystyle\sum_{n}
 (-1)^nb_{n}^{(4)}
\Psi\left(1-i\eta-\frac{B_1}{z_0}-\frac{B_2}{2},
n+\nu+2-B_2-\frac{2B_1}{z_0};2i\omega z\right),\vspace{1.5mm}\\
%
\bar{U}_{4}^{\infty}(z)=e^{i\omega z}(z-z_0)^{1-B_2-\frac{B_1}{z_0}}\times
\vspace{1.5mm}\\
%
\qquad\displaystyle\sum_{n}
c_{n}^{(4)}\Psi\left(n+\nu+1+i\eta-\frac{B_{1}}{z_{0}}-
\frac{B_{2}}{2},n+\nu+2-B_{2}-
\frac{2B_{1}}{z_{0}};-2i\omega z\right),
\end{array}
\end{eqnarray}
where the recurrence relations for $c_{n}^{(4)}$ are
\begin{eqnarray}
&&(n+\nu+1)c_{n+1}^{
(4)} +\bigg[(n+\nu)\left(n+\nu+1-B_2-\frac{2B_1}{z_0}+2i\omega z_{0}\right)
\nonumber\\
%
&&+B_3+\frac{B_1}{z_0}\left(B_2+\frac{B_1}{z_0}-1\right)+i\omega z_{0}\left(2-B_2-\frac{B_{1}}{z_{0}}\right)\bigg]c_{n}^{(4) }\nonumber\\
%
&&+2i\omega z_0 \left(n+\nu+1-B_2-\frac{B_{1}}{z_{0}}\right)
\left(n+\nu+i\eta-\frac{B_2}{2}-
\frac{B_1}{z_0}\right)
c_{n-1}^{(4)}=0.\hspace{1cm}
\end{eqnarray}
The $b_{n}^{(4)}$ and $c_{n}^{(4)}$ are connected  by
\begin{eqnarray}
&c_{n}^{(4)}=\Gamma\big[n+\nu+i\eta+1-
\frac{B_2}{2}-\frac{B_1}{z_0}\big]\ b_{
n}^{(4)}.
\end{eqnarray}

One-sided infinite series result when we take
$\nu=0$ in the two-sided series solutions, since
this restricts the summation to $n\geq 0$.

\section*{3. Whittaker-Ince Limit of the Confluent Heun Equation}

In this section we show that for the Whittaker-Ince limit
of the CHE,  that is, for equation
\antiletra
\begin{eqnarray}\label{ince-che2}
z(z-z_{0})\frac{d^{2}U}{dz^{2}}+(B_{1}+B_{2}z)
\frac{dU}{dz}+
\left[B_{3}+q(z-z_{0})\right]U=0,\quad (q\neq0),
\end{eqnarray}
the solutions of the CHE reduce to expansions in
series of Bessel functions. The procedure is the same
used in  \cite{eu3} for one-sided solutions, but we correct
the following systematic error:  the Bessel functions
of the first kind $J_{\lambda}$ which appear in
the solutions of Ref. \cite{eu3}
must be replaced by $(-1)^nJ_{\lambda}$.

In Sec. 3.1  we write the first set of solutions
for Eq. (\ref{ince-che2}).  We find that the three solutions
have exactly the same series coefficients and no  restriction
must be imposed on the parameters
of Eq. (\ref{ince-che2}) in order to
write one solution as a linear combination
of the others. In Sec. 3.2 we show how these solutions are obtained
from the solutions of the CHE by using the Whittaker-Ince
limit (\ref{ince}).  In Sect. 3.3 we use the transformations
rules for Eq. (\ref{ince-che2}) to generate other
sets of solutions. The limits of the Baber-Hass\'e solutions
are given in \cite{eu3}, but such expansions
are unsuitable to get finite-series
solutions.

\subsection*{3.1. The first set of solutions}

The Whittaker-Ince limit of the fundamental set of solutions given
in Eqs. (\ref{che-primeiro set}) yields ($i=1,2$)
\letra
\begin{eqnarray}\label{solutionInce-1a}
\begin{array}{ll}
U_{{1}}(z)=\displaystyle\sum_{n}
(-1)^ nc_{n}^{(1)}\big(\sqrt{qz}\big)^{-(n+\nu+B_2-1)} J_{n+\nu+B_2-1} \big(2\sqrt{qz}\big),& \forall z,\vspace{1.5mm}\\
%
U_{1}^{(i)}(z)=\displaystyle\sum_{n}
(-1)^ nc_{n}^{(1)}\big(\sqrt{qz}\big)^{-(n+\nu+B_2-1)} H_{n+\nu+B_2-1}^{(i)} \big(2\sqrt{qz}\big),&|z|>|z_{0}|,
\end{array}
\end{eqnarray}
where the limits of the recurrence relations (\ref{che-rec-1a})
and (\ref{che-rec-1b}) are
\begin{eqnarray}\label{solutionInce-1b}
&&(n+\nu+1)c_{n+1}^{(1)}+
\big[(n+\nu)(n+\nu+B_2-1)+B_3\big]c_{n}^{(1)}
\nonumber\\
%
&&+qz_{0}
\big[n+\nu+B_{2}+({B_1}/{z_0})
-1\big]
c_{n-1}^{(1)}=0.
\end{eqnarray}
In these solutions $J_{\lambda}(x)$ denotes Bessel functions
of the first kind of order $\lambda$, whereas
$H_{\lambda}^{(1)}(x)$ and $H_{\lambda}^{(2)}(x)$
denote Hankel functions of first and second kind respectively.
The solution $U_1(z)$ comes from the solution $U_1(z)$
of the CHE, while the solutions denoted by $U_{i}^{(i)}(z)$
follow either from $U_{1}^{(\infty)}(z)$ or
$\bar{U}_{1}^{(\infty)}(z)$. This set admits the Leaver limit $z_0\rightarrow 0$.

The Bessel functions which appear in the solutions
are all independent since their Wronskians are \cite{erdelyi1b}
\begin{eqnarray*}
&&W\big[J_{\lambda}(x),H_{\lambda}^{(1)}(x)\big]={2i}/({\pi x}),
\quad W\big[J_{\lambda}(x),H_{\lambda}^{(2)}(x)\big]=-{2i}/({\pi x}),
\\
&&W\big[H_{\lambda}^{(1)}(x),H_{\lambda}^{(2)}(x)\big]=-{4i}/({\pi x}).\end{eqnarray*}
Then, the relation
\antiletra
\begin{eqnarray}\label{J-H}
J_{\lambda}(x)=\frac{1}{2}\big[H_{\lambda}^{(1)}(x)+
H_{\lambda}^{(2)}(x)\big],
\end{eqnarray}
can be used to write each solution as a linear combination of the others
in a region where the three solutions are valid.

On the other hand, for a fixed $\lambda$ the
asymptotic behaviours of the Bessel functions as
$|x|\rightarrow\infty$ are \cite{erdelyi1b}
\begin{eqnarray}\label{Bessel-assimptotico}
\begin{array}{ll}
J_{\lambda}(x)\sim
\sqrt{\frac{2}{\pi x}}
  \cos\left( x-\frac{1}{2}\lambda\pi-\frac{1}{4}\pi\right),
 &|\arg\ x|<\pi;\vspace{1.5mm}\\
%
H_{\lambda}^{(1)}(x)\sim\sqrt{{2}/({\pi x})}
\ e^{ {i}( x-\frac{1}{2}\lambda\pi-\frac{1}{4}\pi)},
& -\pi<\arg\ x<2\pi;\vspace{1.5mm}\\
%
%
H_{\lambda}^{(2)}(x)\sim\sqrt{{2}/({\pi x})}
\ e^{ -{i}( x-\frac{1}{2}\lambda\pi-\frac{1}{4}\pi)},
& -2\pi<\arg\ x<\pi.
\end{array}
\end{eqnarray}
Thus, for the solutions $U_{1}^{(i)}$
we find
\begin{eqnarray*}
\lim_{z\rightarrow \infty}U_{1}^{(1)}(z)\sim
e^{2i\sqrt{qz}}\ z^{\frac{1}{4}-\frac{B_{2}}{2}-\frac{\nu}{2}}
,\quad\lim_{z\rightarrow \infty}{U}_{1}^{(2)}(z)\sim
e^{- 2i\sqrt{qz}}\  z^{\frac{1}{4}-\frac{B_{2}}{2}-\frac{\nu}{2}}.
\end{eqnarray*}
The behaviour of $U_{1}$ when $z\rightarrow \infty$
is a linear combination of these due to Eq. (\ref{J-H}).

Notice that, by setting
\begin{eqnarray}\label{Mathieu-equation}
\begin{array}{l}
w(u)=U(z), \quad z=\cos^{2}(\sigma u),\quad
( \sigma=1,i),\vspace{1.5mm}\\
%
z_{0}=1,\quad B_{1}=-\frac{1}{2}, \quad B_{2}=1, \quad
B_{3}=\frac{k^2}{2}-\frac{a}{4},\quad q=k^2,
\end{array}
\end{eqnarray}
in Eq. (\ref{ince-che2}), we obtain the
Mathieu equation (\ref{mathieu}). Then, from the solutions
(\ref{solutionInce-1a}) and (\ref{solutionInce-1b})  we
get the following even solutions for the Mathieu
equation
\letra
\begin{eqnarray}\begin{array}{ll}
w_{1}(u)=\displaystyle\sum_{n}
(-1)^ nc_{n}^{(1)}\big[k\cos(\sigma{u})\big]^{-n-\nu} J_{n+\nu} \big(2k\cos(\sigma{u})\big),
& \forall u,\vspace{1.5mm}\\
%
w_{1}^{(i)}(u)=\displaystyle\sum_{n}
(-1)^ nc_{n}^{(1)}\big[k\cos(\sigma{u})\big]^{-n-\nu} H_{n+\nu}^{(i)} \big(2k\cos(\sigma{u})\big),
& |\cos(\sigma{u})|>1,
\end{array}
\end{eqnarray}
where the coefficients  $c_{n}^{(1)}$ satisfy
\begin{equation}
(n+\nu+1)c_{n+1}^{(1)}+
\left[ (n+\nu)^2+({k^2}/{2})-({a}/{4})\right]
c_{n}^{(1)}+k^2 \left[n+\nu-({1}/{2})\right]c_{n-1}^{(1)}=0.
\end{equation}
In this set of two-sided infinity
series solutions, the first solution converges
for any $u$, in contrast with the usual two-sided
solutions for the Mathieu equation which
converge, all of them, only for $|\cos(\sigma{u})|>1$ \cite{abramowitz,meixner}.

One-sided infinite series are obtained by putting
$\nu=0$ in the two-sided series solutions, what restricts
the summation to $n\geq 0$.

\subsection*{3.2. Derivation of the solutions}
To compute the Whittaker-Ince limits, first we rewrite the solutions of
the CHE in a form convenient for using the
formulas (\ref{J}) and (\ref{K}). Thus, we rewrite the
solutions (\ref{che-primeiro set}) as ($q=-2\eta \omega$)
\antiletra\letra
\begin{equation}\label{limite1}
U_{{1}}(z)=e^{-i\omega z}\displaystyle\sum_{n}
\frac{(-1)^nc_{n}^{(1)}}{\Gamma(n+\nu+B_2)}
\ \Phi \left(\frac{B_2}{2}-i\eta,n+\nu+B_2;\frac{qz}{i\eta}\right),
\hspace{1.6cm}
\end{equation}
\begin{equation}\label{limite2}
U_{{1}}^{\infty}(z)=e^{-i\omega z}
\displaystyle\sum_{n}
D_{n} \Gamma\left(1-i\eta-n-\nu-\frac{B_2}{2}\right)
\Psi\left(\frac{B_2}{2}-i\eta,n+\nu+B_2;\frac{qz}{i\eta}\right),
\end{equation}
\begin{eqnarray}\label{limite3}
\bar U_{{1}}^{\infty}(z)=e^{i\omega
z}\displaystyle\sum_{n}
E_{n}(-qz)^{1-n-\nu-B_2}
\Gamma\left(i\eta+n+\nu+\frac{B_2}{2}\right)
\times\hspace{2.3cm}\nonumber\\
%
\Psi\left(1+i\eta-\frac{B_{2}}{2},
2-n-\nu-B_{2};-\frac{qz}{i\eta}\right),
\end{eqnarray}
where the new coefficients are defined by
\begin{eqnarray*}
D_{n}= \frac{(-1)^{n}b_{n}^{(1)}}
{\Gamma\big[1-n-\nu-i\eta-({B_2}/{2})\big]},\quad
E_{n}=\frac{(i \eta)^ {n+\nu}c_{n}^{(1)}}{
\Gamma\big[i\eta+n+\nu+({B_2}/{2})\big]}.
\end{eqnarray*}
Inserting these relations into Eqs. (\ref{che-rec-1a}) and
(\ref{che-rec-1b}) we find
\begin{eqnarray*}
&&(n+\nu+1)D_{n+1}+
\beta_{n}^ {(1)}D_{n}\\
%
&&+2i\omega
z_{0}\big[n+\nu+i\eta+({B_{2}}/{2})-1\big]
\big[n+\nu+B_{2}+({B_{1}}/{z_{0}})
-1\big]D_{n-1}=0,
\end{eqnarray*}
and
\begin{eqnarray*}
&&(n+\nu+1)
\left[ \frac{n+\nu+i\eta+(B_{2}/2)}{i\eta}\right] E_{n+1}+\beta_{n}^ {(1)}E_{n}
\\
%
&&-2\eta\omega
z_{0}\big[n+\nu+B_{2}+({B_{1}}/{z_{0}})
-1\big]
E_{n-1}=0,\hspace{6mm}
\end{eqnarray*}
where
\begin{eqnarray*}
&\beta_{n}^{(1)}=(n+\nu)\left(n+\nu+B_{2}-1+2i\omega z_{0}\right)+
B_{3}+i\omega z_{0}\left( B_{2}+\frac{B_{1}}{z_{0}}\right) .
\end{eqnarray*}
Thence, we find that the Whittaker-Ince limits of these
relations are identical since
\begin{eqnarray*}
\lim c_{n}^{(1)}=\lim D_{n}=\lim E_{n},
\quad(\omega \rightarrow 0,\quad   \eta\rightarrow \infty,\quad
2\eta\omega=-q).
\end{eqnarray*}
Denoting by $c_{n}^{(1)}$ the above limits, we obtain
the recurrence relations (\ref{solutionInce-1b}).

Now, by letting $(-i\eta)\rightarrow \infty$ and
using Eq. (\ref{J}) we find
%
\[\Phi \left(\frac{B_2}{2}-i\eta,n+\nu+B_2;\frac{qz}{i\eta}\right)
\rightarrow\Gamma(n+\nu+B_2)\big(\sqrt{qz}\big)^{1-n-\nu-B_2}
J_{n+\nu+B_{2}-1}\big(2\sqrt{qz}\big).\]
%
Thus the limit of Eq. (\ref{limite1}) is the solution
$U_{1}(z)$ written in Eqs. (\ref{solutionInce-1a}).
On the other hand, to
obtain the limit of the solution (\ref{limite2}) we define
$ L_{1}(z)$ by
\begin{eqnarray*}
&L_{1}(z):=\Gamma\left(1-i\eta-n-\nu-\frac{B_2}{2}\right)
\Psi\left(\frac{B_2}{2}-i\eta,n+\nu+B_2;\frac{qz}{i\eta}\right).
\end{eqnarray*}
Then, when $(-i\eta)\rightarrow\infty$ the relation
(\ref{K}) gives
\begin{eqnarray*}
L_{1}(z)\rightarrow 2\big(\pm i\sqrt{qz}\big)^{1-n-\nu-B_2}
K_{n+\nu+B_2-1}
\big(\pm 2i\sqrt{qz}\big)=\quad\\ \\
\left\{
\begin{array}{l}
-i\pi e^{i\pi (1-n-\nu-B_2)}\big(\sqrt{qz}\big)^{1-n-\nu-B_2}
H_{n+\nu+B_2-1}^{(2)}
\big(2\sqrt{qz}\big),\vspace{1.5mm}\\
i\pi e^{-i\pi (1-n-\nu-B_2)}\big(\sqrt{qz}\big)^{1-n-\nu-B_2}
H_{n+\nu+B_2-1}^{(1)}
\big(2\sqrt{qz}\big),
\end{array}
\right.
\end{eqnarray*}
where in last step we have used the relations among the functions
$K_{\lambda}$ and $H_{\lambda}^{(i)}$ given in
Eq. (\ref{hankel}). Inserting these into the
limit of the solution (\ref{limite2}) and supressing
multiplicative factors,
we find the solutions $U_{1}^{(i)}(z)$
given in (\ref{solutionInce-1a}).
The same solutions follow from the limit of the solution (\ref{limite3}).
Notice that these are  formal derivations which
involve some tricks. However, we may check the resulting
solutions by inserting them into Eq. (\ref{ince-che2}).

%
\subsection*{3.3. Other sets of solutions for Whittaker-Ince Limit of
the CHE}
%
%
%
If $U(z)=U(B_{1},B_{2},B_{3};z_{0},q;z)$ denotes one
solution (or set of solutions) for Whittaker-Ince limit
(\ref{incegswe}) of the CHE, then the rules
$\mathscr{T}_1$, $\mathscr{T}_2$ and $\mathscr{T}_3$
given by \cite{eu2}
\antiletra
\begin{eqnarray}\label{Transformacao3}
\begin{array}{ll}
\mathscr{T}_{1}
U(z)=z^{1+B_{1}/z_{0}}
U(C_{1},C_{2},C_{3};z_{0},q;z),& z_{0}\neq0,\vspace{1.5mm}\\
%
\mathscr{T}_{2}
U(z)=(z-z_{0})^{1-B_{2}-B_{1}/z_{0}}U(B_{1},D_{2},D_{3};
z_{0},q;z), &z_{0}\neq0,\vspace{1.5mm}\\
%
\mathscr{T}_{3}
U(z)=
U(-B_{1}-B_{2}z_{0},B_{2},
B_{3}-q z_{0};z_{0},-q;z_{0}-z),&\forall z_{0}
\end{array}
\end{eqnarray}
can generate a group having 8 solutions. The constants
$C_{i}$ and $D_{i}$ are defined in Eqs. (\ref{constantes-C-D} ).
Next we use only $\mathscr{T}_1$
and $\mathscr{T}_2$, following the sequence
\begin{equation}
\left(U_1,U_{{1}}^{(i)}\right)
\stackrel {\mathscr{T}_1}{\longleftrightarrow}
\left(U_{{2}},U_{{2}}^{(i)}\right)
\stackrel {\mathscr{T}_2}{\longleftrightarrow}
\left(U_{{3}},U_{{3}}^{(i)}\right)
\stackrel {\mathscr{T}_1}{\longleftrightarrow}
\left(U_{{4}},U_{{4}}^{(i)}\right),
%
\end{equation}
where $(U_1,U_{{1}}^{(i)})$ denotes the
set of solutions written in Eqs. (\ref{solutionInce-1a})
and (\ref{solutionInce-1b}).
The conditions for having doubly infinite series are obtained by
choosing $\nu$ such that the coefficients of $c_{n}^{(i)}$ and
$c_{n-1}^{(i)}$ do not vanish in the recurrence
relations, as in the case of the CHE.
Besides this, according to Eq. (\ref{J-H}) one solution can
be expressed as a linear combination of the
other solutions. We will write only the solutions
in series of Bessel functions of the first kind, $J_{\lambda}$;
the others are obtained by replacing $J_{\lambda}$
by the first and second Hankel functions
$H_{\lambda}^{(i)}$ ($i=1,2$). The solutions for the Mathieu
equation are derived by using Eq. (\ref{Mathieu-equation}).

The first solution of  the second set, which does not admit
the Leaver limit $z_0\rightarrow 0$, is ($\forall z$)
\letra
\begin{eqnarray}\label{C2}
U_2(z)=z^{1+\frac{B_{1}}{z_{0}}}\displaystyle
\sum_{n}
(-1)^ nc_{n}^{(2)}\big(\sqrt{qz}\big)^{-n-\nu-1-B_2-\frac{2B_1}{z_0}}
\ J_{n+\nu+1+B_2+\frac{2B_1}{z_0}} \big(2\sqrt{qz}\big),
\end{eqnarray}
where the coefficients satisfy
\begin{eqnarray}
&&(n+\nu+1)c_{n+1}^{(2)}+
\bigg[(n+\nu)\left(n+\nu+B_2+1+\frac{2B_1}{z_0}\right)+B_3\hspace{1cm}\nonumber\\
%
&&+\left(1+\frac{B_1}{
z_0 } \right)
\left(B_2+\frac{B_1}{z_0}\right)\bigg] c_{n}^{(2)} +
qz_0\left(n+\nu+B_2+\frac{B_1}{z_0}-1 \right)
c_{n-1}^{(2)}=0.\hspace{1cm}
\end{eqnarray}
For the Mathieu equation the corresponding solutions are even,
as those of the first set, since
\begin{eqnarray*}
w_{2}(u)=\cos{(\sigma u})\displaystyle\sum_{n}
(-1)^ nc_{n}^{(2)}\big[k\cos(\sigma{u})\big]^{-n-\nu-1}
J_{n+\nu+1} \big(2k\cos(\sigma{u})\big),\quad \forall u,
\end{eqnarray*}
with
\begin{eqnarray*}
&&(n+\nu+1)c_{n+1}^{(2)}+
\left[ (n+\nu)(n+\nu+1)+\frac{k^2}{2}+\frac{1-a}{4}\right]
c_{n}^{(2)}\hspace{2.5cm}\\
%
&&+k^2 \big[n+\nu-({1}/{2})\big]c_{n-1}^{(2)}=0.
\end{eqnarray*}
In third set, which admits the Leaver limit $z_0\rightarrow 0$, the first solution is
\antiletra\letra
\begin{eqnarray}\label{inceCHE-3a}
&&U_3(z)=z^{1+\frac{B_{1}}{z_{0}}}\ (z-z_0)^{1-B_2-\frac{B_1}{z_0}}\times\nonumber\\
%
&&\hspace{1.4cm}\displaystyle
\sum_{n}(-1)^n c_{n}^{(3)}\
\big(\sqrt{qz}\big)^{-n-\nu-3+B_2}\
J_{n+\nu+3-B_2} \big(2\sqrt{qz}\big),\quad
 \forall z\hspace{2cm}
\end{eqnarray}
associated with the recurrence relations
\begin{eqnarray}\label{inceCHE-3b}
&&(n+\nu+1)c_{n+1}^{(3)}+
\left[(n+\nu)(n+\nu+3-B_2)+B_3+2-B_2\right] c_{n}^{(3)}\hspace{2cm} \nonumber\\
%
%
&&+qz_0\big[n+\nu+1-B_2-({B_1}/{z_0}) \big]
c_{n-1}^{(3)}=0.
\end{eqnarray}
For the Mathieu equation, this gives the odd solution:
\begin{eqnarray*}
w_{3}(u)=\sin{(2\sigma u})\displaystyle\sum_{n}
(-1)^n c_{n}^{(3)}\big[k\cos(\sigma{u})\big]^{-n-\nu-2} J_{n+\nu+2}
\big(2k\cos(\sigma{u})\big),\quad \forall u,
\end{eqnarray*}
where
\begin{eqnarray*}
&&(n+\nu+1)c_{n+1}^{(3)}+
\left[ (n+\nu)(n+\nu+2)+\frac{k^2}{2}-\frac{a}{4}+1\right]
c_{n}^{(3)}\hspace{2cm}\\
%
&&+k^2 \big[n+\nu+({1}/{2})\big]c_{n-1}^{(3)}=0.
\end{eqnarray*}

The fourth set of solutions,
which does not admit the Leaver limit, gives
\antiletra\letra
\begin{eqnarray}\label{C-4a}
&&U_4(z)=(z-z_0)^{1-B_2-\frac{B_1}{z_0}}\times\ \nonumber\\
%
&&\hspace{.7cm} \displaystyle\sum_{n}
(-1)^n c_{n}^{(4)}\ \big(\sqrt{qz}\big)^{-n-\nu-1+B_2+\frac{2B_1}{z_0}}
\ J_{n+\nu+1-B_2-\frac{2B_1}{z_0}} \big(2\sqrt{qz}\big),
\ \  \forall z,\hspace{1cm}
\end{eqnarray}
with the recurrence relations
\begin{eqnarray}\label{C-4b}
&&(n+\nu+1)c_{n+1}^{(4)}+
\bigg[(n+\nu)\left(n+\nu+1-B_2-\frac{2B_1}{z_0}\right)+B_3\nonumber\\
%
%
&&+\frac{B_1}{
z_0}\left(B_2+\frac{B_1}{z_0}-1\right)\bigg] c_{n}^{(4)} +
qz_0\left(n+\nu+1-B_2-\frac{B_1}{z_0} \right)
c_{n-1}^{(4)}=0.\hspace{1cm}
\end{eqnarray}
For the Mathieu equation, we find odd solutions:
\begin{eqnarray*}
w_{4}(u)=\sin{(\sigma u})\displaystyle\sum_{n}
(-1)^n c_{n}^{(4)}\big[k\cos(\sigma{u})\big]^{-n-\nu-1} J_{n+\nu+1}
\big(2k\cos(\sigma{u})\big),\quad \forall u,
\end{eqnarray*}
with the recurrence relations
\begin{eqnarray*}
&&(n+\nu+1)c_{n+1}^{(4)}+
\left[ (n+\nu)(n+\nu+1)+\frac{k^2}{2}-\frac{a-1}{4}\right]
c_{n}^{(4)}\hspace{2cm}\\
%
&&+k^2 \big[n+\nu+({1}/{2})\big]c_{n-1}^{(4)}=0.
\end{eqnarray*}

%

%
\section*{4. The Double-Confluent Heun Equation (DCHE)}
In Sec. 4.1 we obtain sets of solutions for the
double-confluent Heun equation (\ref{dche})
by applying the Leaver limit ($z_0\rightarrow 0$)
to solutions of the CHE. The expansions in series
of regular hypergeometric functions converge
for any $z$, whereas the expansions in series of
irregular functions converge for $|z|>0$. We also
give the conditions
to express one solution in terms of the others and
obtain one-sided series solutions by truncating the
two-sided series on the left. Finally, in Sec. 4.2 we
write the solutions
for Eq. (\ref{incedche}), that is, for the Whittaker-Ince
limit of the DCHE.
\subsection*{4.1. Solutions for the general case}
For $z_0\rightarrow 0$ the solutions
(\ref{che-primeiro set}) are not affected formally,
but their recurrence relations change.  We find
\antiletra\letra
\begin{eqnarray}\label{primeiro-dche}
\begin{array}{l}
U_{{1}}(z)=e^{-i\omega z}\displaystyle\sum_{n}
(-1)^{n}b_{n}^{(1)}
\ \widetilde{\Phi}\left(\frac{B_{2}}{2}-i\eta,
n+\nu+B_{2};2i\omega z\right)\vspace{1.5mm}\\
%
U_{{1}}^{\infty}(z)=e^{-i\omega z}\displaystyle\sum_{n}
(-1)^n b_{n}^{(1)}
\Psi\left(\frac{B_{2}}{2}-i\eta,n+\nu+B_{2};2i\omega z\right),\vspace{1.5mm}\\
%
\bar{U}_{{1}}^{\infty}(z)=e^{i\omega z}\displaystyle\sum_{n}
c_{n}^{(1)} \Psi\left(n+\nu+i\eta+\frac{B_{2}}{2},n+\nu+B_{2};-2i\omega z\right).
\end{array}
\end{eqnarray}
The recurrence relations for the
$c_{n}^{(1)}$ are
\begin{eqnarray}
&&(n+\nu+1)c_{n+1}^{(1)}+
\big[ (n+\nu)\left(n+\nu+B_{2}-1\right)+
i\omega B_{1}+B_{3}\big]c_{n}^{(1)}\nonumber\\
%
&&+2i\omega B_{1}\big[n+\nu+i\eta+({B_{2}}/{2})-1\big]
c_{n-1}^{(1)}=0,
\end{eqnarray}
wherefrom we get relations for $b_{n}^{(1)}$
by means of
\begin{eqnarray*}&c_{n}^{(1)}=
\Gamma\left( n+\nu+i\eta+\frac{B_2}{2} \right) b_{n}^{(1)}.
\end{eqnarray*}
The behaviour given in Eq. (\ref{par1-infinito}) for the
CHE is also valid for the present solutions.

These three solutions are given by doubly infinite series if
\antiletra\letra
\begin{eqnarray}
&\nu \text{ and }
\nu+i\eta+\frac{B_{2}}{2} \text{ are not integers.}
\end{eqnarray}
If, besides this,
\begin{eqnarray}
&\nu+B_{2} \text{ is not integer, and }\frac{B_{2}}{2}-i\eta
\text{ is not zero or negative integer,}
\end{eqnarray}
then any of these solutions can be written as a linear
combination of the others by using Eq. (\ref{continuation2}).
Again, as in the CHE, only the restriction on  $(B_{2}/2)-i\eta$ cannot
be satisfied by an suitable choice of $\nu$. In addition, if
$(B_{2}/2)-i\eta=-m$  ($m=0,1,2,\cdots$),
the expansion  $U_{1}$ becomes
meaningless. For this case, instead of $U_{1}$, we can use
\antiletra\letra
\begin{eqnarray}
U_{1}^{\text{p}}(z)=e^{-i\omega z}\displaystyle \sum_{n=0}^{\infty}d_{n}^{(1)}
\left(\frac{z}{B_{1}}\right)^{n},\qquad (|z|=\text{finite})
\end{eqnarray}
where the recurrence relations for the
coefficients are ($d_{-1}^{(1)}=0$)
\begin{eqnarray}
&&(n+1)d_{n+1}^{(1)}+
\left[ n\left(n+B_{2}-1\right)-
i\omega B_{1}+B_{3}\right]d_{n}^{(1)}\nonumber\\
%
&&-2i\omega B_{1}\big[n-i\eta+({B_{2}}/{2})-1\big]
d_{n-1}^{(1)}=0.
\end{eqnarray}
This $U_{1}^{\text{p}}$, which reduces to a Heun
polynomial when $(B_{2}/2)-i\eta=-m$,
was obtained by letting $z_0\rightarrow 0$
in Eq.  (\ref{baber-p}).

A second set follows from the solutions (\ref{B3-a})
and the corresponding recurrence relations (\ref{B3-b}).
Using the L'Hospital rule we find that
\antiletra
\begin{eqnarray}\label{regra de Lopital}
z^{1+\frac{B_{1}}{z_{0}}}
(z-z_{0})^{1-B_{2}-\frac{B_{1}}{z_{0}}}=
z(z-z_{0})^{1-B_{2}}\left(1-\frac{z_{0}}{z}\right)^{-\frac{B_{1}}{z_{0}}}\rightarrow
 z^{2-B_{2}}e^{B_{1}/z},\ (z_{0}\rightarrow 0)
\end{eqnarray}
and, hence, we obtain the solutions
\letra
\begin{eqnarray}\label{DCHE-2}
\begin{array}{l}
U_{2}(z)=z^{2-B_2}e^{-i\omega z+\frac{B_{1}}{z}}
\displaystyle\sum_{n}
(-1)^{n}b_{n}^{(2)}
\widetilde{\Phi}\left(2-i\eta-\frac{B_2}{2},
n+\nu+4-B_2;2i\omega z\right),\vspace{1.5mm}\\

U_{2}^{\infty}(z)=\displaystyle
z^{2-B_2}e^{-i\omega z+\frac{B_{1}}{z}}
\sum_{n}
(-1)^n b_{n}^{(2)}\Psi\left(2-i\eta-\frac{B_2}{2},
n+\nu+4-B_2;2i\omega z\right),\vspace{1.5mm}\\
%
\bar{U}_{2}^{\infty}(z)=z^{2-B_2}e^{i\omega z+\frac{B_{1}}{z}}
\times\vspace{1.5mm}\\
%
\hspace{1.5cm}\displaystyle\sum_{n}
c_{n}^{(2)}\Psi\left(n+\nu+2+i\eta-\frac{B_2}{2},
n+\nu+4-B_2;-2i\omega z\right),
\end{array}
\end{eqnarray}
where the coefficients $c_{n}^{(2)}$ obey
\begin{eqnarray}\label{baber-hasse}
&&(n+\nu+1)c_{n+1}^{(2)}+
\big[(n+\nu)\left(n+\nu+3-B_2\right)+B_3+2-B_2-i\omega B_{1}
\big]c_{n}^{(2)}\nonumber\\
%
&&-2i\omega B_1
\big[n+\nu+i\eta+1-({B_2}/{2})\big]
c_{n-1}^{(2)}=0.\hspace{5.6cm}
\end{eqnarray}
The recurrence relations for the $b_{n}^{(2)}$ are derived
from these via the relation
\[
c_{n}^{(2)}=\Gamma\big[n+\nu+i\eta+2-({B_2}/{2})\big]b_{n}^{(2)}.\]

Now the conditions to assure that the
three solutions are given by doubly infinite series are
\antiletra\letra
\begin{eqnarray}
&\nu \text{ and }
\nu+i\eta-\frac{B_2}{2}\text{ cannot be integers,}
\end{eqnarray}
and if, besides this,
\begin{eqnarray}
&\nu+B_2   \text{ is not integer, and }  i\eta+\frac{B_2}{2}
 \text{ is not zero or negative integer,}
\end{eqnarray}
then Eq. (\ref{continuation2}) can be used
to express one solution in
terms of the others. If $i\eta+(B_2/2)=2,3,\cdots$,
instead of $U_2$ we can use the solution obtained from
(\ref{baber-2}) by the substitution ($\eta,\omega$)$\rightarrow$
($-\eta,-\omega$).

One-sided series solutions are obtained by setting
$\nu=0$ in the two-sided solutions, in the same way as
in Sect. 2.2. We write only the first set and
the limit of the Baber-Hass\'e solution (\ref{baber}).
\antiletra\letra
\begin{eqnarray}\label{dche-truncada}
\begin{array}{l}
U_{1}(z)=e^{-i\omega z}\displaystyle\sum_{n=0}^{\infty}
\frac{(-1)^{n}c_{n}^{(1)}}{\Gamma(n+B_2)}
\ {\Phi}\left(\frac{B_{2}}{2}-i\eta,n+B_{2};2i\omega z\right),\vspace{1.5mm}\\
%
 U_{1}^{\infty}(z)=
e^{-i\omega z}\displaystyle\sum_{n=0}^{\infty}
(-1)^n b_{n}^{(1)}
\ \Psi\left(\frac{B_{2}}{2}-i\eta,n+B_{2};2i\omega z\right),\vspace{1.5mm}\\
%
\bar U_{1}^{\infty}(z)=e^{i\omega z}
\displaystyle\sum_{n=0}^{\infty}
c_{n}^{(1)} \Psi\left(n+i\eta+
\frac{B_{2}}{2},n+B_{2};-2i\omega z\right),
\end{array}
\end{eqnarray}
where the recurrence relations for
$b_{n}^{(1)}$ and $c_{n}^{(1)}$ are
($b_{-1}^{(1)}=c_{-1}^{(1)}=0$)
\begin{eqnarray}
&&(n+1)\big[n+i\eta+({B_{2}}/{2})\big]b_{n+1}^{(1)}+
\big[ n\left(n+B_{2}-1\right)+i\omega B_{1}+B_{3}\big]\ b_{n}^{(1)}\hspace{1cm}
\nonumber\\
%
&&+2i\omega B_{1}\
b_{n-1}^{(1)}=0,
\end{eqnarray}
\begin{eqnarray}
&&(n+1)c_{n+1}^{(1)}+
\left[ n\left(n+B_{2}-1\right)+
i\omega B_{1}+B_{3}\right]c_{n}^{(1)}\nonumber\\
%
&&+2i\omega B_{1}\big[n+i\eta+({B_{2}}/{2})-1\big]
c_{n-1}^{(1)}=0.
\end{eqnarray}
The limit of the Baber-Hass\'e solution (\ref{baber}) is
\antiletra
\begin{eqnarray}
U_{1}^{\text{baber}}(z)=e^{i\omega z}\displaystyle \sum_{n=0}^{\infty}c_{n}^{(1)}
\left(\frac{z}{B_{1}}\right)^{n}.
\end{eqnarray}
Notice that the solution $U_{1}^{\infty}(z)$ does not
admit finite-series solutions, while the other solutions do.
However, for finite-series it is sufficient to consider
only Baber-Hass\'e solution. On the other hand,
the condition in order that the three expansions in
series of confluent hypergeometric functions are
given by infinite series ($n\geq 0$) is
\begin{eqnarray}
&i\eta+\frac{B_2}{2}
\text{ is not zero or negative integer,}
\end{eqnarray}
since in this case the series do not truncate on
the left or on right. In addition, if
\begin{eqnarray}
&B_{2}\text{ and  }\frac{B_{2}}{2}-i\eta
\text{ are not zero or negative integer,}
\end{eqnarray}
then Eq. (\ref{continuation2}) can be used to
connect the three one-sided infinite-series solutions.

The Baber-Hass\'e solution corresponding to
the second set is \cite{eu3}
\begin{eqnarray}\label{baber-2}
U_{2}^{\text{baber}}(z)=e^{i\omega z+(B_{1}/z)}z^{2-B_{2}}
\displaystyle \sum_{n=0}^{\infty}c_{n}^{(2)}
\left(-\frac{z}{B_{1}}\right)^{n},
\end{eqnarray}
where the recurrence relations for $c_{n}^{(2)}$
are obtained by putting $\nu=0$ in Eq. (\ref{baber-hasse}).
The corresponding solutions in series of
hypergeometric functions can
be obtained by taking $\nu=0$
in the solutions (\ref{DCHE-2}).
%

%
%
\subsection*{4.2. Solutions for the Whittaker-Ince limit
of the DCHE}
For the Whittaker-Ince limit (\ref{incedche}) of DCHE,
namely,
\begin{eqnarray*}
z^2\frac{d^{2}U}{dz^{2}}+(B_{1}+B_{2}z)
\frac{dU}{dz}+
\left(B_{3}+qz\right)U=0,\qquad (q\neq0, \ B_{1}\neq 0),
\end{eqnarray*}
there is no  finite-series solutions. For this reason,
the equation is irrelevant for quasi-exactly solvable problems.
However, its solutions  may be important
for studying the scattering of ions which induce
dipole and quadrupole moments in polarisable
targets \cite{eu2}.
Solutions for the above equation can be found
by applying either the Whittaker-Ince
limit to the solutions given in  Sec. 4.1 or the Leaver limit
to the solutions (\ref{solutionInce-1a}) and (\ref{inceCHE-3a}).
We use the latter approach because it is the easiest.

In effect the Leaver limit $z_0\rightarrow 0$ does not modify
the solutions (\ref{solutionInce-1a}), which again read
\letra
\begin{eqnarray}\begin{array}{ll}
U_{{1}}(z)=\displaystyle\sum_{n}
(-1)^ nc_{n}^{(1)}\big(\sqrt{qz}\big)^{-(n+\nu+B_2-1)} J_{n+\nu+B_2-1}
\big(2\sqrt{qz}\big),& \forall z,\vspace{1.5mm}\\
%
U_{1}^{(i)}(z)=\displaystyle\sum_{n}
(-1)^ nc_{n}^{(1)}\big(\sqrt{qz}\big)^{-(n+\nu+B_2-1)} H_{n+\nu+B_2-1}^{(i)}
\big(2\sqrt{qz}\big),&|z|>0,
\end{array}
\end{eqnarray}
but changes their recurrence relations to
\begin{eqnarray}
(n+\nu+1)c_{n+1}^{(1)}+
\big[(n+\nu)(n+\nu+B_2-1)+B_3\big]c_{n}^{(1)}+qB_1c_{n-1}^{(1)}=0.
\end{eqnarray}
These three solutions are really doubly infinite if $\nu$ is
not integer. Furthermore, each solution
can be expressed as a linear combination of the
others by means of the relation (\ref{J-H}).

On the other side, the only detail to get the Leaver limit
of the solutions given in Eqs. (\ref{inceCHE-3a}) is the use
the limit written in Eq. (\ref{regra de Lopital}).
The results are
\antiletra\letra
\begin{eqnarray}\begin{array}{l}
U_2(z)=e^{\frac{B_{1}}{z}} z^{2-B_2}\displaystyle
\sum_{n}(-1)^n c_{n}^{(2)}\
\big(\sqrt{qz}\big)^{-n-\nu-3+B_2}\
J_{n+\nu+3-B_2} \big(2\sqrt{qz}\big),\quad \forall z;\vspace{1.5mm}\\
%
U_2^{(i)}(z)=e^{\frac{B_{1}}{z}} z^{2-B_2}\displaystyle
\sum_{n}
(-1)^ nc_{n}^{(2)}
\big(\sqrt{qz}\big)^{-n-\nu-3+B_2}
H_{n+\nu+3-B_2}^{(i)} \big(2\sqrt{qz}\big),\ |z|>0
\end{array}
\end{eqnarray}
where
\begin{equation}
n+\nu+1)c_{n+1}^{(2)}+
\left[(n+\nu)(n+\nu+3-B_2)+B_3+2-B_2\right] c_{n}^{(2)}
-qB_1c_{n-1}^{(2)}=0.
\end{equation}

One-sided series are obtained by taking $\nu=0$ once more.

%
%
\section*{5. Conclusions}

In this section we collect some results we have found,
mention other solutions which deserve
further analysis, and indicate the possibility of applying
the solutions of the four equations given in the first section to
physical problems.

We have started with a
set of solutions for the confluent Heun equation (CHE)
given by three series
of confluent hypergeometric functions and,
by means of Leaver and the Whittaker-Ince limits,
we have derived sets of solutions
for all the equations included in the diagram
described in the first section. The Baber-Hass\'e
expansions in power series also admit the both
limits and can be used to provide solutions
for the cases in which the solutions in
hypergeometric functions are not valid, as well as
to get finite-series solutions for the confluent and
double-confluent Heun equations.

In the fundamental set of two-sided solutions
($ U_{1},U_{1}^{\infty}, \bar{U}_{1}^{\infty}$),
the first and the second solutions are, respectively, Leaver's
expansions in series of regular and irregular confluent
hypergeometric functions for the CHE \cite{leaver1},
while $\bar{U}_{1}^{\infty}$
is the two-sided version of the
expansion in series of irregular
functions given in Ref. \cite{eu3}. We have seen that,
although the solution  $U_{1}$ is not valid if there is
a certain constraint between two parameters of the equation,
we can use the transformation rules to find a two-sided
solution valid for that case. We have also
established the conditions
to express each of the three
solutions as a linear combination of the others.

The fact that the two-sided expansion $U_{1}$
in series of regular functions converges
for any $z$, and in particular in a neighbourhood of
$z=0$,   distinguishes
the present solutions from the Leaver
expansions in series of Coulomb
wave functions \cite{leaver1}, since the
latter converge only for $|z|>|z_{0}|$. On the other hand,
we have seen that
\begin{eqnarray*}
&\displaystyle\lim_{z\rightarrow \infty}U_{1}^{\infty}(z)
\sim e^{-i\omega z}z^{i\eta-\frac{B_2}{2}}
,\qquad\lim_{z\rightarrow \infty}\bar{U}_{1}^{\infty}(z)
\sim e^{i\omega z}z^{-i\eta-\nu-\frac{B_2}{2}},
\end{eqnarray*}
where $\nu$ is the characteristic parameter
of the two-sided series. The behaviour of $U_{1}$ when
$z\rightarrow\infty$ is given by a
combination of these. This explains
why $U_1$ is sometimes
inappropriate when the variable $z$ tends to infinity.
%

All the solutions for the double-confluent Heun
equation (DCHE), obtained from solutions of the CHE
through the Leaver limit ($z_0\rightarrow{0}$),
are also given by expansions in series of
confluent hypergeometric functions. Then, the analysis
used for the solutions of the CHE has been
promptly adapted to the solutions of the DCHE,
including the conditions which allow to write
one solution of a fixed set as a linear
combination of the others.

We have also found one-sided infinite-series solutions
for the CHE and DCHE, by truncating
the two-side infinite series on the left. The one-sided
solutions for the CHE can afford two different types
of finite-series solutions. One of these may be ignored
because they are also suplied by the Baber-Hass\'e expansions
in power series. The other type of finite series
is important because it is the only one preserved by
the Whittaker-Ince limit of the CHE.

The Whittakker-Ince limit (\ref{ince}) of
the CHE and DCHE  have generated
expansions in series of Bessel functions
of the first kind and in series of the first and second
Hankel functions.  In this case, each solution belonging
to fixed set can be written
as a linear combination of the others without the need of
restrictions on the parameters of the differential
equations. The Whittakker-Ince limit of
the DCHE does not admit finite-series solutions.

In order to have a full account of the solutions, we
have considered the transformations rules
which result from substitutions of
variables which transform a Heun equation into another
form of itself  (up to a change of parameters).
For instance, only if we apply the transformation
rules to solutions
of the CHE, can we derive solutions
for the Mathieu and Whittaker-Hill equations having
all the desired properties of parity and periodicity.
Similarly, if an expansion is not valid for certain
values of the parameters,
the transformation rules allow to find another
solution valid for that case.

Solutions in series of Coulomb wave
functions \cite{leaver1} and in series
of Gauss hypergeometric functions
\cite{eu1,mano1,mano2} for the confluent Heun equation
also admit both the Leaver and the Whittaker-Ince limits
\cite{eu1,eu2}.
However, there are several
unexplored aspects concerning such expansions.
Thus, it is necessary to consider other solutions
of this type in order to retrieve the usual solutions
of the ordinary spheroidal equation
\cite{abramowitz,arscott,meixner}, which is a particular
case of Eq. (\ref{gswe}) with $\eta=0$ \cite{leaver1}.
This would amount new solutions for the equations
of the diagram written in the first section.
On the other side, by means of the Whittaker-Ince limit
these solutions conduct
to known solutions for the Mathieu equations \cite{eu2}.
Nevertheless, the possibility of finding solutions
for the CHE which reproduce all the  known solutions of
the Mathieu equations is an open issue.

Now we refer to some problems which obey the CHE and its
limiting cases. In the first place, the Schr\"odinger
equation with inverse fourth and sixth-power
potentials leads, respectively,  to the DCHE (\ref{dche})
and its Whittaker-Ince limit (\ref{incedche}).
These potentials have appeared, for example, in the description
of intermolecular forces \cite{frank} and in the
scattering of ions by polarisable atoms \cite{kleinman}.

In fact, the radial equation for the scattering of ions  by
neutral targets
with induced dipole and quadrupole moments
can be put in the form of the Whittaker-Ince limit
of the DCHE \cite{eu2};
for neutral or charged targets presenting only
induced dipole moment, the equation may be reduced to the
DCHE \cite{buhring,eu2}. In these problems, there is no
arbitrary constant in the differential equations and so,
in order to assure the series convergence,
it is necessary to use two-sided infinite series solutions
with a characteristic parameter $\nu$.

In the second place, as already mentioned, the equations
which rule the time-dependence
of Klein-Gordon and Dirac test-fields in  some nonflat
Friedmannian spacetimes can be written as CHEs and DCHEs
\cite{birrell,eu1}. These equations also require two-sided
infinite series solutions. In addition, there are the CHEs
for angular and radial Teukolsky equations in Kerr spacetimes
\cite{bini,leaver1,mano1,mano2}.

Finally, in Appendix A we have
found that normal forms of the
Heun equations are helpful to know
if a given one-dimensional Schr\"odinger
equation reduces to Heun equations. In
this connection, we have seen that there are
quasi-exactly solvable (QES)
potentials representative of each of the five Heun equations,
a fact which may become important if we
have sufficient knowledge about the solutions of
the equation.

As noticed at the beginning, QES problems
demand finite-series solutions which give only a
part of the energy spectrum, but in the case of Heun equations
there are also infinite-series solutions. Whether these
afford the remaining part of the spectrum,
this is a theme to be analysed. At last, we
observe that the inverted QES potential (\ref{cho})
leads to the Whittaker-Ince limit of the
confluent Heun equation. As far as we are aware,
the only other example known for this equation
(excepting Mathieu equations)
comes from the separation of variables of the
Laplace-Beltrami operator for scalar
fields in an Eguchi-Hanson space
\cite{birkandan,malmendier,mignemi}.

\section*{Appendix A. Heun Equations and
Quasiexact Solvability}
\protect\label{A}
\setcounter{equation}{0}
\renewcommand{\theequation}{A.\arabic{equation}}

We consider the Heun equations in their
normal or Schr\"odinger form,
\begin{eqnarray}
\label{normal}
\left[ \frac{d^2}{dz^2}+Q(z)\right] y(z)=0,
\end{eqnarray}
that is, in the form where there is no first-order
derivative term \cite{decarreau1}.
The functionÃÂ  $Q(z)$ for the general Heun equation
and its confluent cases is given in the following
\cite{ronveaux}, where it is understood
that, in general, there exists
some constraint among the parametres $A$, $B$, and so on.
In each case we
indicate the singular points of the equation.

\indent
\textit{General Heun equation}. Four regular singular points
at $z=0,1,a,\infty$:
\begin{equation}\label{geral}
Q(z)=\frac{A}{z}+\frac{B}{z-1}+\frac{C}{z-a}+\frac{D}{z^2}+
\frac{E}{(z-1)^2}+\frac{F}{(z-a)^2}, \qquad(a\neq 0\  \text{or }\ 1).
\end{equation}
\indent
\textit{Confluent Heun equation} (or generalised spheroidal
wave equation). Two regular points at $z=0,1$ and one irregular
at $z=\infty$:
\begin{eqnarray}\label{confluente}
Q(z)=A+\frac{B}{z}+\frac{C}{z-1}+\frac{D}{z^2}+
\frac{E}{(z-1)^2}.
\end{eqnarray}
\indent
\textit{Double-confluent Heun equation}. Two irregular points at
$z=0,\infty$:
\begin{eqnarray}\label{dupla}
Q(z)=A+\frac{B}{z}+\frac{C}{z^2}+\frac{D}{z^3}+\frac{E}{z^4}.
\end{eqnarray}
\indent
\textit{Biconfluent Heun equation}. One regular point at $z=0$
and one irregular point at $z=\infty$:
\begin{eqnarray}\label{bi}
Q(z)=Az^2+Bz+C+\frac{D}{z}+
\frac{E}{z^2}.
\end{eqnarray}
\indent
\textit{Triconfluent Heun equation}. One irregular point at $z=\infty$:
\begin{eqnarray}\label{tri}
Q(z)=Az^4+Bz^3+Cz^2+Dz+E.
\end{eqnarray}

Other normal forms are
given in the tables
constructed by Lemieux and Bose \cite{lemieux} who, however,
have not considered the triconfluent equation. These tables
are helpful to recognise whether a given
equation is of the Heun type. In particular, by using the
Lemieux-Bose tables in addition to
the normal forms written above, it is straightforward
to establish relations among the Heun and
the Schr\"odinger equations for some
quasi-exactly solvable potentials.

We find: (i) a triconfluent Heun equation
for the quartic potential  given in Eq. (5.34)
of Gonz\'alez-L\'opez, Kamran and Olver \cite{gonz};
(ii) biconfluent Heun equations
for the sextic potential $V_{1}z^6+V_{2}z^4+V_{3}z^2
+V_{4}+ V_{5}/z^2$ given by Turbiner
\cite{turbiner1} and Ushveridze \cite{ushveridze2}, and
for the potentials II, III and VIII
given in Turbiner's list \cite{turbiner1};
(iii) double-confluent Heun equations for
the inverse fourth-power potential
$V(r)=V_1r^{-4}+V_2r^ {-3}+V_{3}r^ {-2}+V_4r^{-1}$
 \cite{turbiner,turbiner1}, and
for the asymmetric double-Morse potential
given by Zaslavskii and Ulyanov \cite{zaslavskii}; (iv)
confluent Heun equations for the trigonometric and
hyperbolic potentials given by Ushveridze \cite{ushveridze1};
(v) general Heun equations in
the Darboux elliptic form \cite{darboux} for the first and second
Ushveridze's elliptic potentials \cite{ushveridze1}.

Finally we write a potential given by Cho and Ho \cite{cho},
namely,
\begin{eqnarray}\label{cho}
V(u)=-\frac{\text{b}^{2}}{4}\sinh^{2}u-\left(\ell^{2}-
\frac{1}{4}\right)\frac{1}{\cosh^{2}u},\quad u\in(-\infty,
\infty),\quad (\ell=1,2,3,\cdots)
\end{eqnarray}
where $\text{b}$ is a positive real constant. This is a bottomless potential
in the sense that $V(u)\rightarrow{-}\infty$
when $u\rightarrow\pm \infty$. If $\text{b}^2<4\ell^2-1$,
it is an inverted double-well potential; if
$\text{b}^2\geq4\ell^2-1$, the potential is similar to the
one of an inverted oscillator.
For this potential the Schr\"{o}dinger
reduces to the Whittaker-Ince
limit (\ref{incegswe}) of the confluent Heun equation.
%
%
\section*{Appendix B. Confluent
Hypergeometric Functions}
\protect\label{B}
\setcounter{equation}{0}
\renewcommand{\theequation}{B.\arabic{equation}}
The regular and irregular confluent
hypergeometric functions, denoted respectively by $\Phi(a,c;y)$
and $\Psi(a,c;y)$, satisfy the Kummer transformations
\begin{eqnarray}\label{kummer}
\Phi(\mathrm{a,c};y)=e^{y}\Phi(c-a,c;-y),
\qquad \Psi(a,c;y)=y^{1-c}\Psi(1+a-c,2-c;y).
\end{eqnarray}
The behaviour of ÃÂ of $\Psi(a,c;y)$ and
$\Phi(a,c;y)$ when $y\to \infty$ is given by
\begin{eqnarray}
\label{asymptotic-confluent1}
\lim_{y\rightarrow \infty}\Psi(a,c;y)\sim y^{-a}[1+O(|y|^{-1})],
\qquad ÃÂ \left(-\frac{3\pi}{2}< \arg{y}<\frac{3\pi}{2}\right)
\end{eqnarray}
\begin{eqnarray}\label{asymptotic-confluent2}
\lim_{y\rightarrow \infty}\Phi(a,c;y)=\left\{
\begin{array}{ll}\displaystyle
\frac{\Gamma(c)}
{\Gamma(a)}e^{y}y^{a-c}
[1+O(|y|^{-1})], &(\Re{y}>0)
\vspace{1.5mm}\\
\displaystyle\frac{\Gamma(c)}
{\Gamma(c-a)}(-y)^{-a}[1+O(|y|^{-1})], &(\Re{y}<0).
\end{array}
\right.
\end{eqnarray}
These can be used to get the limits of the series expansions
as $y\to\infty$. For
$\Re{y}=0$, the limit of $\Phi(a,c;y)$ is a combination
of the limits on the right-hand side of the previous
expression \cite{erdelyi1}.

To find the expansions in series of irregular confluent
hypergeometric
functions for the CHE, in addition to Eq. (\ref{confluent0}), we
need the relations
\begin{eqnarray}\label{relation2}
\begin{array}{l}
y\frac{d\Psi(a,c;y)}{dy}=(1-c)\Psi(a,c;y)+(c-a-1)\Psi(a,c-1;y),\vspace{1.5mm}\\
%
\frac{d\Psi(a,c;y)}{dy}=\Psi(a,c;y)-\Psi(a,c+1;y),\vspace{1.5mm}\\
%
(c-a-1)
\Psi(a,c-1;y)+(1-c-y)\Psi(a,c;y)+y\Psi(a,c+1;y)=0,
\end{array}
\end{eqnarray}
where the last relation results from the first and second ones.
These relations also hold for the  functions $\tilde{\Phi}(a,b;y)$ defined in
Eq. (\ref{phi-tilde}). On the other hand, if $c\to\infty$, while $a$
and $y$ remain bounded,  we have \cite{erdelyi1}
\begin{eqnarray*}
&&\Psi(a,c;y)=(c)^{-a}\left[
(-1)^{-a} +\big({\sqrt{2\pi}}/{\Gamma(a)}\big)
\ c^{c+a-(3/2)}y^{1-c}e^{y-c}\right]\left[1+O(|c|^{-1})\right]\\
%
&&= (c)^{-a}\left[
(-1)^{-a}+\frac{\sqrt{2\pi}}{\Gamma(a)}
\ \left(\frac{c}{ey} \right) ^{c+a-({3}/{2})}y^{a-(1/2)}e^{y+a-(3/2)}\right]
 \left[1+O\left( \frac{1}{|c|}\right) \right]
\end{eqnarray*}
which, in conjunction with the last realion of Eq. (\ref{relation2}), is useful
in the study of the convergence of the series solutions for the CHE.
Notice that this equation implies
\begin{eqnarray}
\lim_{c\rightarrow-\infty}\Psi(a,c;y) \sim (-c)^{-a}
\big[1+O(|c|^{-1})\big],\hspace{5.2cm}\\
%
\lim_{c\rightarrow+\infty}\Psi(a,c;y) \sim
\frac{\sqrt{2\pi}}{\Gamma(a)}
\ \left(\frac{c}{ey} \right) ^{c+a-\frac{3}{2}}c^{-a}y^{a-\frac{1}{2}}e^{y+a-(3/2)}
 \left[1+O\left( \frac{1}{|c|}\right) \right].
\end{eqnarray}

The Whittaker-Ince limit of the expansions in series of
confluent hypergeometric functions for the CHE and DCHE
yields expansions in series of Bessel functions for
Eqs. (\ref{incegswe}) and (\ref{incedche}) by means of
the limits  \cite{erdelyi1}
\begin{eqnarray}\label{J}
\displaystyle
\lim_{a\rightarrow ÃÂ \infty}
\Phi\left(a,c;-\frac{y}{a}\right)=
\Gamma(c)\ y^{(1-c)/2}J_{c-1}\big(2\sqrt{y}\big),\hspace{5.45cm}
\end{eqnarray}
\begin{eqnarray}\label{K}
\displaystyle
\lim_{a\rightarrow ÃÂ \infty}\left[\Gamma(a+1-c)\ \Psi\left(a,c;
\frac{y}{a}\right)\right]=
2y^{(1-c)/2}K_{c-1}\big(2\sqrt{y}\big),\hspace{3.8cm}
\end{eqnarray}
\begin{eqnarray}\label{H}
\displaystyle
\lim_{a\rightarrow ÃÂ \infty}\left[\Gamma(a+1-c)\ \Psi\left(a,c;
-\frac{y}{a}\right)\right]=
\begin{cases}\displaystyle
-i\pi e^{i\pi c}y^{(1-c)/2}H_{c-1}^{(1)}\big(2\sqrt{y}\big),
 \text{Im}\ y>0,%
\vspace{4mm}\\
\displaystyle
i\pi e^{-i\pi c}y^{(1-c)/2}H_{c-1}^{(2)}\big(2\sqrt{y}\big),
\text{Im}\ y<0,%
\end{cases}
\end{eqnarray}
where $J_{\lambda}$ denotes the
Bessel functions of the first kind of order $\lambda$,
$K_{\lambda}$ are modified Bessel
functions of the third kind,
whereas $H_{\lambda}^{(1)}$ and $H_{\lambda}^{(2)}$
denote the first and second Hankel functions \cite{erdelyi1b},
respectively. Some relations useful for Sec. 3.2 are
\begin{eqnarray}\label{hankel}
\begin{array}{ll}
H_{-\lambda}^{(1)}(x)=e^{i\pi\lambda}H_{\lambda}^{(1)}(x),
&
H_{-\lambda}^{(2)}(x)=e^{-i\pi\lambda}H_{\lambda}^{(2)}(x),\vspace{1.5mm}\\
%
K_{\lambda}(-ix)=\frac{1}{2}\pi ie^{\frac{1}{2}\pi i\lambda}
H_{\lambda}^{(1)}(x),& K_{\lambda}(ix)=-\frac{1}{2}
\pi ie^{-\frac{1}{2}\pi i\lambda}H_{\lambda}^{(2)}(x).
\end{array}
\end{eqnarray}
%

%

\end{document}